\documentclass[3p,12pt]{elsarticle}
\usepackage[utf8]{inputenc}
\usepackage{amsmath}
\usepackage{amssymb}
\usepackage{amsfonts}
\usepackage[numbers]{natbib}
\usepackage{booktabs}
\usepackage{paralist}
\usepackage{listings}
\usepackage{footnote}
\usepackage{float}
\usepackage{hyperref}
\usepackage{color}
\usepackage{url}
\usepackage{graphicx}
\usepackage[parfill]{parskip}

\hypersetup{
  colorlinks=true,
  urlcolor=blue,
  linkcolor=blue,
  citecolor=black,
  filecolor=black
}

\makeatletter
\def\ps@pprintTitle{%
  \let\@oddhead\@empty
  \let\@evenhead\@empty
  \def\@oddfoot{\footnotesize\itshape\hfill\today}%
  \let\@evenfoot\@oddfoot}
\makeatother

\newcounter{labelscope}
\newcommand\newlabelscope{\refstepcounter{labelscope}}

\newcommand\labellst[1]{lst:\thelabelscope:#1}
\newcommand\reflst[1]{\ref{lst:\thelabelscope:#1}}

\newcommand\repolink[1]{\href{https://github.com/jdmota/tools-examples/tree/8649e64da4787234345ded3d836c5ef87724ecde/#1}{github.com/.../#1}}
\newcommand\codelink[2]{\href{https://github.com/jdmota/tools-examples/tree/8649e64da4787234345ded3d836c5ef87724ecde/#1}{#2}}

\begin{document}

\begin{frontmatter}

\title{On using VeriFast, VerCors, Plural, and KeY to check object usage}

\author[NOVA]{João Mota}
\author[NOVA]{Marco Giunti}
\author[NOVA]{António Ravara}

\address[NOVA]{NOVA LINCS and NOVA School of Science and Technology, Portugal}

\begin{abstract}
Typestates are a notion of behavioral types that describe protocols for stateful objects, specifying the available methods for each state, in terms of a state machine. Usually, objects with protocol are either forced to be used in a linear way, which restricts what a programmer can do, or deductive verification is required to verify programs where these objects may be aliased. To evaluate the strengths and limitations of static verification tools for object-oriented languages in checking the correct use of shared objects with protocol, we present a survey on four tools for Java: VeriFast, VerCors, Plural, and KeY. We describe the implementation of a file reader and of a linked-list, check for each tool its ability to statically guarantee protocol compliance as well as protocol completion, even when objects are shared in collections, and evaluate the programmer's effort in making the code acceptable to these tools.
\end{abstract}

\begin{keyword}
Java \sep Typestates \sep VeriFast \sep VerCors \sep Plural \sep KeY
\end{keyword}

\end{frontmatter}

\tableofcontents

\section{Introduction}

When programming in an object-oriented language, one naturally defines objects where their methods' availability depends on their internal state~\cite{nierstrasz1993regular,ancona2016behavioral}.
One might represent their intended usage protocol with an automaton or a state machine~\cite{DBLP:journals/corr/abs-2009-08769,TypestateEditor,DBLP:conf/sblp/DuarteR21}.
\textbf{Behavioral types} allow us to statically check if all code of a program respects the protocol of each object. In session types approaches~\cite{DBLP:conf/esop/HondaVK98,DBLP:journals/csur/HuttelLVCCDMPRT16}, objects associated with protocols are usually forced to be used in a linear way to avoid race conditions, which reduces concurrency and restricts what a programmer can do.
Given that \textbf{sharing of objects} is very common, it should be supported. For example, pointer-based data structures, such as linked-lists, usually rely on internal sharing (i.e. aliasing). Such collections may also be used to store objects with protocol and state which needs to be tracked.
Moreover, it is crucial that all \textbf{protocols complete} to ensure responsiveness, for instance, that required method calls are not forgotten, and that all resources are freed.

To study the contributions and limitations of the state of the art, with regards to the static verification of programs with sharing of objects, we present a survey on four tools for Java: VeriFast~\cite{DBLP:conf/aplas/JacobsSP10,jacobs2011verifast}, VerCors~\cite{DBLP:conf/isola/HuismanM20,DBLP:conf/ifm/BlomDHO17}, Plural~\cite{bierhoff2007modular}, and KeY~\cite{DBLP:series/lncs/10001}. We picked these tools because they provide rich features for verification, which we highlight next, and because they are actively maintained (with the exception of Plural). VeriFast is a modular verifier for programs annotated with method contracts written in \textbf{separation logic}~\cite{o2001local,reynolds2002separation}. It comes with a dedicated IDE which allows one to observe each step of a proof when an error is encountered. VerCors is a modular checker which employs \textbf{permission-based concurrent separation logic}~\cite{o2007resources} to check programs, inspired in Chalice~\cite{leino2009basis,leino2009verification}. Both VeriFast and VerCors support \textbf{fractional permissions}~\cite{bornat2005permission}, and use Z3~\cite{de2008z3}, an SMT solver, in the backend. Plural is a plugin for the Eclipse IDE which verifies that the protocols of objects are respected with an approach based on \textbf{typestates}~\cite{StromY86}. It also introduces the notion of \textbf{access permissions} which combine typestate and object aliasing information, allowing state to be tracked and modified even when objects are shared. A comprehensive survey on access permission-based specifications was presented by~\citet{sadiq2020survey}. KeY is a interactive theorem prover for sequential Java programs based on first-order Java dynamic logic (JavaDL)~\cite{harel1984dynamic} with support for specifications written in JML~\cite{DBLP:journals/sigsoft/LeavensBR06}.

VeriFast and VerCors are similar in that both employ separation logic and often require \textbf{deductive reasoning}~\cite{DBLP:journals/expert/BeckertH14}, via the definition of lemmas, to be able to verify more complex programs, and insertion of assertions in key program points to guide verification. VerCors has support for permission-based specifications, inspired in Chalice~\cite{leino2009basis,leino2009verification}, while VeriFast does not. Plural applies a type system~\cite{pfenning2004benjamin} directly supporting typestates and different types of access permissions, allowing for more kinds of object sharing, beyond the ``single writer vs multiple readers'' model of fractional permissions. Although it is not maintained any longer, one can install it by downloading its source\footnote{\url{https://code.google.com/archive/p/pluralism/}}, the source code of its dependencies, and installing them in Eclipse Juno\footnote{\url{https://www.eclipse.org/downloads/packages/release/juno}} (an old version from 2012). Given its support for rich access permissions, and direct application of the typestate abstraction, we believe its study is still relevant. KeY distinguishes itself from the aforementioned tools in that it supports the greatest number of Java features, verification is based on first-order dynamic logic, which is expressive enough to encode Hoare logic~\cite{hoare1969axiomatic}, and provides an interactive theorem prover with a high degree of automation and useful tactics which the programmer can use to guide proofs.

For each tool we: (1) assess if it can check the \textbf{correct use of objects with protocol}, including \textbf{protocol completion}, even with objects \textbf{shared in collections}; (2) evaluate the \textbf{programmer's effort} in making the code acceptable to it.
To these ends, we present implementations of file readers with a usage protocol and linked-list collections. In the cases we succeed in implementing the linked-list, we also develop an iterator for such collection. Then we show examples of code using these objects. All code is available online.\footnote{\repolink{}} We believe attempting to implement a linked-list is particularly relevant because it is a common data structure. Furthermore, its use of pointers often creates challenges for less expressive type systems, so it is a great candidate for use case examples.\footnote{For example, in Rust, one has to follow an ownership discipline, preventing one from creating linked-lists, unless \textit{unsafe} code is used. GhostCell~\cite{DBLP:journals/pacmpl/YanovskiDJD21}, a recent solution to deal with this, allows for internal sharing but the collection itself still needs to respect the ownership discipline. GhostCell uses \textit{unsafe} code for its implementation but was proven safe with separation logic.}

\begin{figure}
  \centering
  \includegraphics[width=0.5\linewidth]{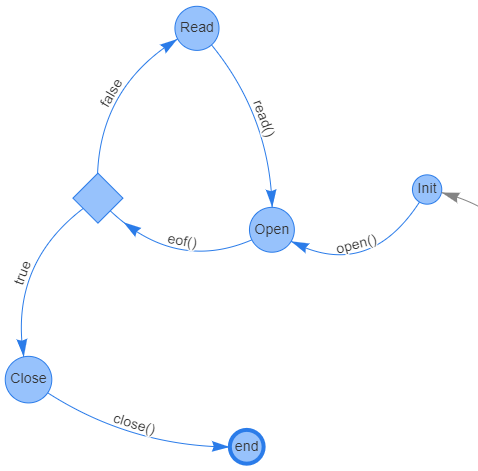}
  \caption{File reader's protocol}
  \label{fig:filereaderprotocol}
\end{figure}

The file reader has a usage protocol (Figure~\ref{fig:filereaderprotocol}) such that one must first call the {\it open} method, followed by any number of {\it read} calls, until the end of the file is reached (checked by calling the {\it eof} method), and then terminated with the {\it close} method. The linked-list is single-linked, meaning that each node has a reference only to the next node. This list has two fields: {\it head} and {\it tail}. The former points to the first node, the latter points to the last node, as it is commonly implemented in imperative languages. Items are added to the {\it tail} and removed from the {\it head}, following a FIFO discipline. The implementations are in Java, a language supported by all the aforementioned tools. In particular, given its object-oriented nature, Java is well-suited for building objects with protocol where method calls act like transitions of a state-machine.

This survey is structured as follows:

\begin{itemize}
\item Section~\ref{sec:background} presents \textbf{background information} on each tool;
\item Section~\ref{sec:results} discusses our \textbf{results} together with \textbf{detailed assessments}.
\item Section~\ref{sec:conclusions} presents our \textbf{conclusions}.
\end{itemize}

\section{Background}
\label{sec:background}

In this section, we provide overviews of each tool we study in this survey.

\newlabelscope

\subsection{VeriFast}

VeriFast~\cite{DBLP:conf/aplas/JacobsSP10,jacobs2011verifast} is a modular verifier for (subsets of) C and Java programs annotated with method contracts (pre- and post-conditions)~\cite{meyer1992applying} written in \textbf{separation logic}~\cite{o2001local,reynolds2002separation}. Besides the points-to assertions from separation logic, specifications support the definition of inductive data types, predicates, fixpoint definitions and lemma functions. VeriFast is then able to statically check that contracts are respected during execution and that programs will not raise errors such as null pointer exceptions or perform incorrect actions such as accessing illegal memory. Nonetheless, VeriFast's support for generics is still limited.\footnote{https://github.com/verifast/verifast/issues/271}

For the sharing of memory locations, VeriFast has built-in support for \textbf{fractional permissions}~\cite{bornat2005permission}, associating a number coefficient between 0 (exclusive) and 1 (inclusive) to each heap chunk. By default, the coefficient is 1, which allows reads from and writes to a memory location. A number less than 1 allows only for reads. The programmer may provide coefficient patterns in the form of expressions, such as literal numbers or variables, or in the form of existentially quantified names, as exemplified in List.~\reflst{coefficientexample}. These patterns may be applied to points-to assertions but also to predicates. Applying a coefficient to a predicate is equivalent to multiplying it by each coefficient of each heap chunk mentioned in the predicate's body. Additionally, VeriFast supports the automatic splitting and merging of fractional permissions. Counting permissions~\cite{bornat2005permission} are also supported via a trusted library.% (i.e. they are not built-in).

\begin{minipage}{\linewidth}
\begin{lstlisting}[language=Java, caption=Coefficient pattern example, label=\labellst{coefficientexample}]
int getBalance()
  //@ requires [?f]account(this, _);
  //@ ensures [f]account(this, _);
{
  return this.balance;
}
\end{lstlisting}
\end{minipage}

For C programs, VeriFast also supports leak checking: after consuming post-conditions, the heap must be empty. However, the programmer can indicate that leaking a certain resource is acceptable with the {\it leak a;} command. For Java programs, leaking is always allowed.

\newlabelscope

\subsection{VerCors}

VerCors~\cite{DBLP:conf/isola/HuismanM20,DBLP:conf/ifm/BlomDHO17} is a verifier for concurrent programs written in Java, C, OpenCL and PVL (Prototypal Verification Language), and annotated with method contracts. The specifications employ a logic based on \textbf{permission-based concurrent separation logic}~\cite{o2007resources} which also uses \textbf{fractional permissions}~\cite{bornat2005permission} to track sharing. The verification procedure is modular, checking each method in isolation given a contract with pre- and post-conditions. Although there are other tools that perform static verification on annotated programs, VerCors focuses on supporting different concurrency patterns of high-level languages, and is designed to be language-independent. Support for inheritance and exceptions it is still being worked upon based on theoretical work by~\citet{essay81338}.\footnote{https://vercors.ewi.utwente.nl/wiki/\#inheritance-1} Internally, VerCors uses the Viper backend~\cite{DBLP:conf/vmcai/0001SS16}, which in turn uses Z3~\cite{de2008z3}.

\begin{minipage}{\linewidth}
\begin{lstlisting}[language=Java, caption=Permissions example, label=\labellst{permissionsexample}]
/*@ 
  requires Perm(val, 1);
  ensures  Perm(val, 1) ** val == \old(val) + 1;
@*/
void increment(){
 val = val + 1;
}
\end{lstlisting}
\end{minipage}

VerCors supports two styles for specifying access to memory locations: \textbf{permission annotations}, following the approach of Chalice~\cite{leino2009basis,leino2009verification}; and \textbf{points-to assertions} of separation logic, as in VeriFast~\cite{DBLP:conf/aplas/JacobsSP10,jacobs2011verifast}. Both styles of specification have been shown to be equivalent by~\citet{DBLP:conf/esop/ParkinsonS11}. The equivalence is presented in Figure~\ref{permissions_vs_points_to}. On the left-hand-side it is shown the use of a permission annotation, \texttt{Perm}, to request access to variable \texttt{var}, with fractional permission \texttt{p}, and storing a value equal to \texttt{val}. The \texttt{**} symbol represents the separating conjunction operator. On the right-hand-side it is shown the equivalent \texttt{PointsTo} assertion. Permission annotations are very useful because they allow us to refer to values in variables without the need to use new names for them.

\begin{figure}[h]
\centering
$Perm(var, p)\ **\ var == val \equiv PointsTo(var, p, val)$
\caption{Permission annotations equivalent to points-to assertions}
\label{permissions_vs_points_to}
\end{figure}

\newlabelscope

\subsection{Plural}

\citet{bierhoff2005lightweight} addressed the problem of {\bf substitutability of subtypes} while guaranteeing {\bf behavioral subtyping} in a object-oriented language. The specification technique models protocols using abstract states, incorporating {\bf state refinements} (allowing the definition of substates, thus supporting substitutability of subtypes), {\bf state dimensions} (which define orthogonal states corresponding to AND-states in Statecharts~\cite{Harel87}), and {\bf method refinements} (allowing methods in subclasses to accept more inputs and return more specific results). The approach is similar to pre- and post-condition based ones but provides better information hiding thanks to the typestate~\cite{StromY86} abstraction. %Union and intersection types are used as the underlying semantics of specifications, guaranteeing behavioral subtyping and adding expressiveness.

%Since the purpose was to explore the usefulness of powerful specifications (which are currently difficult to check statically), the approach does not impose aliasing restrictions. Instead, the authors implemented a dynamic analysis to check Java programs which monitors method executions with AspectJ.\footnote{\url{https://www.eclipse.org/aspectj/}} Java annotations are used to provide the specifications.

%To deal with dangling resources, \citet{bierhoff2005lightweight} propose an approach where the root state, called {\it alive}, is refined into two states, {\it collect} and {\it bound}. In the former, the object is available for garbage collection, while in the latter it is not. The dynamic analysis then checks all objects that become available for garbage collection for potential dangling resources.

%Unfortunately, the technique treats arrays as normal objects and does not consider their length. Additionally, it does not deal with changes of control flow caused by exceptions. Protocol completion is also not discussed.

\citet{bierhoff2007modular} then built on previous work~\cite{bierhoff2005lightweight} and developed a sound modular protocol checking approach, based on typestates, to ensure at compile-time that clients follow the usage protocols of objects even in the presence of aliasing. For that, they developed the notion of \textbf{access permissions} which combine typestate and object aliasing information. The approach was realized in Plural, a static verifier they developed for Java. As far as we know, not all Java's features are supported, such as exceptions.

\begin{minipage}{\linewidth}
\begin{lstlisting}[language=Java, caption=Iterator example, label=\labellst{iteratorexample}]
@Refine({
  @States(value={"available", "end"}, refined="alive")
})
interface Iterator<E> {
  @Pure("alive")
  @TrueIndicates("available")
  @FalseIndicates("end")
  boolean hasNext();

  @Full(requires="available", ensures="alive")
  E next();
}
\end{lstlisting}
\end{minipage}

An \textbf{access permission} tracks how a reference is allowed to read and/or modify the referenced object, how the object might be accessed through other references, and what is currently known about the object's typestate. To increase the precision of access permissions, \citet{bierhoff2007modular} also introduced \textbf{weak permissions} (such as \textit{share} and \textit{pure}), where an object can be modified through other permissions. The proposed access permissions include a {\bf state guarantee} which ensures that an object remains in that state even in the face of possible changes through other references. Additionally, they track {\bf temporary state assumptions} which are discarded when they become outdated. 
Since access permissions represent resources that need to be consumed (and not duplicated), specifications are based on linear logic~\cite{girard1987linear}. All kinds of permissions are described in Table~\ref{permission_kinds_in_plural}.
%Pre- and post-conditions are separated with a linear implication and use conjunction and disjunction.
%Additionally, one can specify separate permissions to different {\it dimensions}~\cite{bierhoff2005lightweight} of the same object.

\begin{table}[h]
\caption{Kinds of permission in Plural}
\centering
\begin{tabular}{ l | l | l }
Kind & Access to the referenced object & Access other aliases may have\\
\hline
Full & read and write & read-only\\
Pure & read-only & read and write\\
Immutable & read-only & read-only\\
Unique & read and write & none\\
Shared & read and write & read and write\\
\end{tabular}
\label{permission_kinds_in_plural}
\end{table}

These kinds of permissions may be split to allow sharing of an object, and joined back together to allow one to potentially restore {\it unique} permission. {\bf Fractional permissions}~\cite{boyland2003checking} are used to track how much a permission was split. Furthermore, different fractions can be mapped to different state guarantees through a {\bf fraction function}, thus tracking for each state guarantee separately how many other permissions rely on it.

%To deal with inheritance, the approach is based on Fugue's frames~\cite{deline2004typestates}: each object is composed of multiple frames, one for each class in the object's class hierarchy, where each one holds the fields defined in the corresponding class. The frame corresponding to the object's runtime type is called {\it virtual frame}. Unlike Fugue, this approach allows subclasses to explicitly express their expectations of the super-frame's state, thus decoupling typestates of different frames. Additionally, permissions for different frames are distinguished and so, permissions for other frames are only accessible from inside a subclass through \textit{super}. Note that field accesses and assignments are syntactically restricted to the receiver class. Explicit getter and setter methods should be defined to give others access to fields.

%\citet{bierhoff2007modular} also refined the notion of {\bf unpacking}~\cite{deline2004typestates} to allow access to fields of an object. To avoid inconsistencies, specially in the presence of reentrant calls, objects are always fully packed~\cite{deline2004typestates} when methods are called. This is not a limitation because one can pack to some intermediate state.

\citet{DBLP:conf/oopsla/BeckmanBA08} extended the approach to verify the correctness of usage protocols in concurrent programs, statically preventing races on the abstract state of an object as well as preventing violations of state invariants. This approach uses atomic blocks and was also realized in Plural. In this solution, access permissions are used as an approximation of the thread-sharedness of objects. For example, if \textit{pure} or \textit{share} permissions are used, it means that other references can modify the object, and it is assumed that this includes concurrent modifications. In this scenario, temporary state assumptions are discarded, unless the access is synchronized. Furthermore, accessing fields of an object with \textit{share}, \textit{pure}, or \textit{full} permissions, must be performed inside atomic blocks.
%To track which expressions occur in an atomic block, a simple type and effect system~\cite{DBLP:conf/popl/MooreG08} is employed. Expressions typed with the \textit{wt} effect are known to occur in a transaction, expressions typed with \textit{ot} are known to occur outside a transaction, and the \textit{emp} effect marks whose expressions for which the system is not sure~\cite{DBLP:conf/oopsla/BeckmanBA08}.

\citet{DBLP:conf/oopsla/Beckman09} later presented a similar approach which uses synchronization blocks instead of atomic blocks as the mutual exclusion primitive, given that the former are in more wide use. Since programmers are required to synchronize on the receiver object, it becomes implicit to the analysis which parts of the memory are exclusively available, meaning that programmers are not required to specify which parts of the memory are protected by which locks. However, this also implies that private objects cannot be used for the purposes of mutual exclusion.

\newlabelscope

\subsection{KeY}

KeY~\cite{DBLP:series/lncs/10001} is a verifier for sequential Java programs. Specifications are provided in Java comments in an extension of the Java Modeling Language (JML)~\cite{DBLP:journals/sigsoft/LeavensBR06}, called JML*. JML is based on the design by contract~\cite{meyer1992applying} paradigm with class invariants and method contracts. Class invariants describe properties of an object's state that must be preserved by all methods. Method contracts are composed mainly by pre-conditions and post-conditions.
KeY employs modular verification meaning that each method is verified against its contract alone independently from other methods. For this to work, it is also important that contracts specify frame conditions, indicating the heap locations which a method may modify.

\begin{minipage}{\linewidth}
\begin{lstlisting}[language=Java, caption=JML specification example, label=\labellst{jmlexample}]
/*@ normal_behavior
  @ assignable footprint;
  @ ensures getX() == value;
  @ ensures \new_elems_fresh(footprint); 
  @*/
void setX(int value) {
  x = value;
}
\end{lstlisting}
\end{minipage}

As far as we know, KeY is the verification tool that supports the greatest number of Java features, allowing one to verify real programs considering the actual Java runtime semantics, which includes reasoning about inheritance, dynamic method lookup, runtime exceptions, and static initialization. A consequence of this is that, as the members of the KeY team point out~\cite{DBLP:journals/sttt/BrunsMU15}, KeY is not overly suitable for the verification of algorithms that require abstracting away from the code since KeY's main goal is the verification of Java programs.

KeY's core is based on an interactive theorem prover for first-order Java dynamic logic (JavaDL)~\cite{harel1984dynamic}, which can be seen as a generalization of Hoare logic~\cite{hoare1969axiomatic}. An important part in the construction of proofs in KeY is symbolic execution. This process takes every possible execution branch and transforms the program leading to a set of constraints, which can then be verified against the specification. KeY provides a semi-automated proof environment where the user may choose to apply every step of the proof, apply a strategy macro, combining several deductive steps, or execute an automated proof search strategy. Not only KeY offers a high degree of automation, but it also supports SMT solvers, such as Z3~\cite{de2008z3}, which are often useful to solve arithmetical problems~\cite{DBLP:journals/sttt/BrunsMU15}. More details on how to use KeY may be found online.\footnote{\url{https://www.key-project.org/docs/UsingKeyBook/}}

The most common strategy macros are listed below:

\begin{itemize}
  \item Propositional expansion (without splits): apply only non-splitting propositional rules;
  \item Propositional expansion (with splits): apply only propositional rules;
  \item Finish symbolic execution: apply only rules for modal operators of dynamic logic (thus executing Java programs symbolically);
  \item Close provable goals: automatically close all goals that are provable, but do not apply any rules to goals which cannot be closed.
\end{itemize}

\section{Results}
\label{sec:results}

In this section, we present the results of the experiments with each one of the tools and the corresponding detailed assessments.

\newlabelscope

\subsection{VeriFast}

\subsubsection*{File reader implementation}

To model the file reader's protocol, we use pre- and post-conditions in all public methods indicating the expected state and the destination state after the call. To require access to the object's fields and keep track of the current state, we define the {\it filereader} predicate (lines 1-4 of List.~\reflst{filereaderclass1}). The only input parameter is the reference to the file reader. The output parameters (after the semicolon) are the \textit{state} and \textit{remaining} values. Output parameters need to be precisely defined in the predicate and allow the user of the predicate to ``extract'' them. The \textit{state} field is an integer value that tracks the current state. We use constants to identify different states, thus avoiding the use of literal numbers in specifications (lines 7-9). The \textit{remaining} field is the number of bytes left to read, which should always be greater than or equal to zero. When it is 0, it means we reached the end of the file. Note that the separating conjunction binary operator is represented by the \texttt{\&*\&} symbol in VeriFast.

The use of the \textit{state} field may be seen as unnatural and superfluous, but in the absence of a primitive notion of typestate, it is unavoidable. One example of this encoding of protocols being employed is in the Casino contract presented in the VerifyThis event.\footnote{\url{https://verifythis.github.io/casino/}}

\begin{minipage}{\linewidth}
\begin{lstlisting}[language=Java, caption={\it FileReader} class (part 1), label=\labellst{filereaderclass1}]
/*@
predicate filereader(FileReader file; int s, int r) =
  file.state |-> s &*& file.remaining |-> r &*& r >= 0;
@*/

public class FileReader {
  public static final int STATE_INIT = 1;
  public static final int STATE_OPENED = 2;
  public static final int STATE_CLOSED = 3;
\end{lstlisting}
\end{minipage}

Consider, for example, the \textit{eof} method (List.~\reflst{filereaderclass2}). Its contract specifies that we require access to a file reader in the \textit{Opened} state, with a given \textit{remaining} value, named \textit{r} (line 39). The \textit{?} symbol functions like an existential operator and ``extracts'' the last output parameter. Then we ensure that the file remains in the same state with the same \textit{remaining} value, and that the boolean result is \textit{true} if and only if \textit{r} is zero (line 40). The rest of the implementation is very straightforward.\footnote{\repolink{verifast/basic/FileReader.java}}

\begin{minipage}{\linewidth}
\begin{lstlisting}[language=Java, caption={\it FileReader} class (part 2), label=\labellst{filereaderclass2}, firstnumber=38]
  public boolean eof()
    //@ requires filereader(this, STATE_OPENED, ?r);
    //@ ensures filereader(this, STATE_OPENED, r) &*& (result == (r == 0));
  {
    return this.remaining == 0;
  }
  ...
}
\end{lstlisting}
\end{minipage}

\subsubsection*{Linked-list implementation}

The linked-list implementation\footnote{\repolink{verifast/basic/LinkedList.java}} is adapted and extended from a C implementation available online\footnote{\url{https://people.cs.kuleuven.be/~bart.jacobs/verifast/examples/iter.c.html}}. One key difference from the aforementioned C code is that when the linked-list is empty, the {\it head} and {\it tail} fields have {\it null} values, instead of pointing to a dummy node. This matches common implementations and makes verification more challenging because we have to avoid null pointer errors.

To model the linked-list we use the {\it llist} predicate (List.~\reflst{llistpredicate}). This predicate holds the heap chunks of the {\it head} and {\it tail} fields and of all the nodes in the list. The only input parameter is the reference to the linked-list. The output parameters are the references to the head and tail, and a ghost list to allow us to reason about the values in the list in an abstract way. We expose the head and tail for practical reasons when implementing the iterator, detailed in the following section. Lines 3 and 4 ensure that if one of the fields is \textit{null}, the other is also \textit{null} and the list is empty. To ease the addition of new elements to the list, which requires easy access to the tail node, we request access to the sequence of nodes between {\it head} (inclusive) and {\it tail} (exclusive), through the {\it lseg} predicate (List.~\reflst{lsegpredicate}), and then keep access to the {\it tail} directly in the body of the {\it llist} predicate (line 5). The \textit{node} predicate holds the permissions for the \textit{next} and \textit{value} fields of a given node. Note that we do not hold permission to the fields of the values stored. This is to allow them to change independently of the linked-list.

\begin{minipage}{\linewidth}
\begin{lstlisting}[language=, caption={\it llist} predicate, label=\labellst{llistpredicate}]
predicate llist(LinkedList javalist; Node h, Node t, list<FileReader> list) =
  javalist.head |-> h &*& javalist.tail |-> t &*&
    h == null ? t == null &*& list == nil :
    t == null ? h == null &*& list == nil :
    lseg(h, t, ?l) &*& node(t, null, ?value) &*&
      list == append(l, cons(value, nil)) &*& list != nil;
\end{lstlisting}
\end{minipage}

One may notice that we refer to the \textit{FileReader} type directly instead of making the \textit{LinkedList} generic over types of elements. The reason for this is that when we tried to specify a pre-condition (or post-condition) such as \texttt{llist<T>(this, \_, \_, ?list)}, VeriFast would report that ``No such type parameter, inductive datatype, class, interface, or function type:  T'', even though the type parameter was defined in the outer class. There are workarounds to implement a generic linked-list, like using \textit{Object} instead of a type parameter and keeping the information about the actual type around. However, we believe that would be cumbersome. So, we decide to stick with \textit{FileReader} for the purposes of our example. As Bart Jacobs points out, at the time of writing, ``support for Java generics in VeriFast is in its infancy''.\footnote{\url{https://github.com/verifast/verifast/issues/271\#issuecomment-1134814121}}

The {\it lseg} predicate (List.~\reflst{lsegpredicate}) holds the heap chunks of the nodes in the sequence starting on {\it n1} (inclusive) and ending on {\it n2} (exclusive), mapping to the elements of the last and only output parameter {\it list}.

\begin{minipage}{\linewidth}
\begin{lstlisting}[language=, caption={\it lseg} predicate, label=\labellst{lsegpredicate}]
predicate lseg(Node n1, Node n2; list<FileReader> list) =
  n1 == n2 ?
    list == nil :
    node(n1, ?next, ?value) &*& lseg(next, n2, ?l) &*& list == cons(value, l);
\end{lstlisting}
\end{minipage}

The implementation of the {\it remove}\footnote{\repolink{verifast/basic/LinkedList.java\#L60-L78}} and {\it isEmpty}\footnote{\repolink{verifast/basic/LinkedList.java\#L80-L87}} methods is straightforward requiring only the opening and closing of the {\it llist} and {\it lseg} predicates a few times. The implementation of the {\it add}\footnote{\repolink{verifast/basic/LinkedList.java\#L42-L58}} method requires a lemma (List.~\reflst{addlemma}) which states that if we have a sequence of nodes plus the final node, and we append another node in the end, we get a new sequence with all the nodes from the previous sequence, the previous final node, and then the newly appended node.

\begin{minipage}{\linewidth}
\begin{lstlisting}[language=, caption={\it add\_lemma} lemma, label=\labellst{addlemma}]
lemma void add_lemma(Node n1, Node n2, Node n3)
  requires lseg(n1, n2, ?l) &*& node(n2, n3, ?value) &*& node(n3, ?n4, ?v);
  ensures lseg(n1, n3, append(l, cons(value, nil))) &*& node(n3, n4, v);
\end{lstlisting}
\end{minipage}

The \textit{add} method (List.~\reflst{addmethod}) accepts a file reader to be added (line 1). Its contract requires that we have permission for the state of the linked-list, and specifies that the current list of values will be called \textit{list} (line 2). Then the method ensures that the new list is built from the old one by appending the new added value (line 3).

\begin{minipage}{\linewidth}
\begin{lstlisting}[language=Java, caption={\it add} method, label=\labellst{addmethod}]
public void add(FileReader x)
  //@ requires llist(this, _, _, ?list);
  //@ ensures llist(this, _, _, append(list, cons(x, nil)));
{
  ...
}
\end{lstlisting}
\end{minipage}

\subsubsection*{Iterator implementation}

To model the iterator\footnote{\repolink{verifast/basic/LinkedListIterator.java}} we use the {\it iterator} predicate (List.~\reflst{iteratorpredicate}) which holds access to the current node in the {\it curr} field and all the nodes in the linked-list. The last two output parameters are the list of elements already read and the list of elements still to read, respectively. With the {\it iterator\_base} predicate (List.~\reflst{iteratorbasepredicate}), the permissions to the nodes are split in two parts. Half of the permissions ``preserves the structure'' of the list (line 4), and the other half holds the view of the iterator (line 5): a sequence of nodes from the {\it head} (inclusive) to the current node (exclusive), using the {\it lseg} predicate (List.~\reflst{lsegpredicate}); and a sequence from the current node (inclusive) to the final one, using the {\it nodes} predicate (List.~\reflst{nodespredicate}).

\begin{minipage}{\linewidth}
\begin{lstlisting}[language=, caption={\it iterator} predicate, label=\labellst{iteratorpredicate}]
predicate iterator(LinkedList javalist, LinkedListIterator it, Node n;
  list<FileReader> list, list<FileReader> a, list<FileReader> b) =
  it.curr |-> n &*& iterator_base(javalist, n, list, a, b);
\end{lstlisting}
\end{minipage}

\begin{minipage}{\linewidth}
\begin{lstlisting}[language=, caption={\it iterator\_base} predicate, label=\labellst{iteratorbasepredicate}]
predicate iterator_base(LinkedList javalist, Node n;
  list<FileReader> list, list<FileReader> a, list<FileReader> b) =
  [1/2]javalist.head |-> ?h &*& [1/2]javalist.tail |-> ?t &*&
  [1/2]llist(javalist, h, t, list) &*&
  [1/2]lseg(h, n, a) &*& [1/2]nodes(n, b) &*& list == append(a, b);
\end{lstlisting}
\end{minipage}

\begin{minipage}{\linewidth}
\begin{lstlisting}[language=, caption={\it nodes} predicate, label=\labellst{nodespredicate}]
predicate nodes(Node n; list<FileReader> list) =
  n == null ?
    list == nil :
    node(n, ?next, ?value) &*& nodes(next, ?l) &*& list == cons(value, l);
\end{lstlisting}
\end{minipage}

To create the iterator\footnote{\repolink{verifast/basic/LinkedListIterator.java\#L19-L26}}, we take the permission to all the nodes in the linked-list and split it in the aforementioned way. After iterating through all the nodes, the full permission to the nodes needs to be restored to the list. These actions are done with the {\it prepare\_iterator} (List.~\reflst{prepareiteratorlemma}) and the {\it dispose\_iterator} (List.~\reflst{disposeiteratorlemma}) lemmas, respectively. These require some other auxiliary lemmas, namely one showing that the result of appending a list with an element is not an empty list\footnote{\repolink{verifast/basic/Lemmas.java\#L27-L35}}.

\begin{minipage}{\linewidth}
\begin{lstlisting}[language=, caption={\it prepare\_iterator} lemma, label=\labellst{prepareiteratorlemma}]
lemma void prepare_iterator(LinkedList javalist)
  requires llist(javalist, ?h, ?t, ?list);
  ensures iterator_base(javalist, h, list, nil, list);
\end{lstlisting}
\end{minipage}

\begin{minipage}{\linewidth}
\begin{lstlisting}[language=, caption={\it dispose\_iterator} lemma, label=\labellst{disposeiteratorlemma}]
lemma void dispose_iterator(LinkedListIterator it)
  requires iterator(?javalist, it, null, ?list, list, nil);
  ensures llist(javalist, _, _, list);
\end{lstlisting}
\end{minipage}

The implementation of the {\it hasNext} method\footnote{\repolink{verifast/basic/LinkedListIterator.java\#L28-L38}} is straightforward and requires only the opening and closing of some predicates. The implementation of the {\it next} method\footnote{\repolink{verifast/basic/LinkedListIterator.java\#L40-L59}} requires opening and closing predicates, the use of a lemma showing that the {\it append} function is associative (result already available in VeriFast), and the {\it iterator\_advance} lemma, which helps us advance the state of the iterator, moving the just retrieved value from the ``to see'' list to the ``seen'' list (List.~\reflst{iteratoradvancelemma}).

\begin{minipage}{\linewidth}
\begin{lstlisting}[language=, caption={\it iterator\_advance} lemma, label=\labellst{iteratoradvancelemma}]
lemma void iterator_advance(Node h, Node n, Node t)
  requires [1/2]lseg(h, n, ?a) &*& [1/2]node(n, ?next, ?val1) &*&
    [1/2]nodes(next, ?b) &*& [1/2]lseg(h, t, ?list) &*& [1/2]node(t, null, ?val2);
  ensures [1/2]lseg(h, next, append(a, cons(val1, nil))) &*&
    [1/2]nodes(next, b) &*& [1/2]lseg(h, t, list) &*& [1/2]node(t, null, val2);
\end{lstlisting}
\end{minipage}

\subsubsection*{Use example}

To exemplify the use of a linked-list with file readers, we instantiate a linked-list and add three file readers in the {\it Init} state to it (lines 1-7 of List.~\reflst{main}). Then we pass the list to the {\it useFiles} method (line 12) which iterates through all the files and executes their protocol to the end. To assert a fact about elements of the list we use the {\it foreachp} predicate provided by VeriFast, which automatically \textit{closes} the invariant predicates for each file reader. However, we still need to explicitly \textit{close} the {\it foreachp} predicate itself a few times (lines 9-11). The assertion in line 8 ``extracts'' the current list of values and binds it to the name \textit{l}.

\begin{minipage}{\linewidth}
\begin{lstlisting}[language=Java, caption={\it Main} code, label=\labellst{main}]
LinkedList list = new LinkedList();
FileReader f1 = new FileReader("a");
FileReader f2 = new FileReader("b");
FileReader f3 = new FileReader("c");
list.add(f1);
list.add(f2);
list.add(f3);
//@ assert llist(list, _, _, ?l);
//@ close foreachp(cons(f3, nil), INV(FileReader.STATE_INIT));
//@ close foreachp(cons(f2, cons(f3, nil)), INV(FileReader.STATE_INIT));
//@ close foreachp(l, INV(FileReader.STATE_INIT));
useFiles(list);
\end{lstlisting}
\end{minipage}

The pre- and post-conditions of the \textit{useFiles} method ensure that we get a list with all file readers in the {\it Init} state and we end up with the same list but with the readers in the {\it Closed} state (List.~\reflst{usefilescontract}). This is possible because the predicate which models the iterator keeps track of the elements already seen and the elements to see.

\begin{minipage}{\linewidth}
\begin{lstlisting}[language=Java, caption={\it useFiles} contract, label=\labellst{usefilescontract}]
public static void useFiles(LinkedList list)
  //@ requires llist(list, _, _, ?l) &*& foreachp(l, INV(FileReader.STATE_INIT));
  //@ ensures llist(list, _, _, l) &*& foreachp(l, INV(FileReader.STATE_CLOSED));
\end{lstlisting}
\end{minipage}

The verification of the {\it useFiles} method (List.~\reflst{usefilescode}) requires loop invariants to be provided (lines 3-6 and line 13), the opening and closing of the {\it foreachp} predicate a few times (lines 9 and 19-20), the {\it foreachp\_append} lemma provided by VeriFast (which says that if something is true of all elements of two lists, it is also true of the concatenation of the two) (line 21), and the {\it dispose\_iterator} lemma (List.~\reflst{disposeiteratorlemma}) (which restores the access of the nodes to the linked-list) (line 23).

\begin{minipage}{\linewidth}
\begin{lstlisting}[language=Java, caption={\it useFiles} code, label=\labellst{usefilescode}]
LinkedListIterator it = list.iterator();
while (it.hasNext())
  /*@ invariant it != null &*& iterator(list, it, _, l, ?a, ?b) &*&
      foreachp(a, INV(FileReader.STATE_CLOSED)) &*&
      foreachp(b, INV(FileReader.STATE_INIT));
    @*/
{
  FileReader f = it.next();
  //@ open foreachp(b, _);
  //@ open INV(FileReader.STATE_INIT)(f);
  f.open();
  while (!f.eof())
    //@ invariant f != null &*& filereader(f, FileReader.STATE_OPENED, _);
  {
    //@ open filereader(f, _, _);
    f.read();
  }
  f.close();
  //@ close foreachp(nil, INV(FileReader.STATE_CLOSED));
  //@ close foreachp(cons(f, nil), INV(FileReader.STATE_CLOSED));
  //@ foreachp_append(a, cons(f, nil));
}
//@ dispose_iterator(it);
\end{lstlisting}
\end{minipage}

\subsubsection*{Protocol completion}

In an attempt to ensure that all file readers created through the lifetime of the program reach the end of their protocol, we define the \textit{tracker} predicate in the \textit{tracker.javaspec} file\footnote{\repolink{verifast/rt/tracker.javaspec}} which keeps hold of the number of open file readers. This proposed solution is based on a private exchange with Jacobs~\cite{private_conversation_bart_jacobs}. Then, we augment the file reader's specification\footnote{\repolink{verifast/protocol-completion-2/FileReader.java}} to increment this counter in the constructor (List.~\reflst{filereaderconstructor2}), and decrement the counter in the \textit{close} method (List.~\reflst{filereaderclose2}). To effectively increment or decrement the counter, we use the \textit{increment\_tracker} and \textit{decrement\_tracker} lemmas in the implementation of the constructor and of the \textit{close} method. Finally, we assert in the pre-condition and post-condition of \textit{main} that the counter should be 0 (List.~\reflst{trackermain2})\footnote{\repolink{verifast/protocol-completion-2/Main.java\#L8-L9}}.

\begin{minipage}{\linewidth}
\begin{lstlisting}[language=Java, caption=FileReader's constructor, label=\labellst{filereaderconstructor2}]
public FileReader(String filename)
  //@ requires tracker(?n);
  //@ ensures tracker(n + 1) &*& filereader(this, STATE_INIT, _);
{
  this.state = STATE_INIT;
  this.remaining = 20;
  //@ increment_tracker();
}
\end{lstlisting}
\end{minipage}

\begin{minipage}{\linewidth}
\begin{lstlisting}[language=Java, caption=FileReader's \textit{close} method, label=\labellst{filereaderclose2}]
public void close()
  //@ requires tracker(?n) &*& filereader(this, STATE_OPENED, 0);
  //@ ensures tracker(n - 1) &*& filereader(this, STATE_CLOSED, 0);
{
  this.state = STATE_CLOSED;
  //@ decrement_tracker();
}
\end{lstlisting}
\end{minipage}

\begin{minipage}{\linewidth}
\begin{lstlisting}[language=Java, caption={\it main} method, label=\labellst{trackermain2}]
public static void main(String[] args)
  //@ requires tracker(0);
  //@ ensures tracker(0);
\end{lstlisting}
\end{minipage}

Unfortunately, it is possible to fail to ensure protocol completion if the programmer is not careful. Firstly, one could forget to increment and decrement the counter when the typestated-object is initialized and when its protocol finishes, respectively. Secondly, if one forgets to add the post-condition to the {\it main} method, protocol completion will not be actually enforced. So, we can guarantee protocol completion but only if the programmer does not fall for these ``traps''. Here we see that ghost code is useful to check some properties, but if such code is not correctly connected with the ``real'' code, then the property we desired to establish is not actually guaranteed.

An alternative solution is to use a ghost list that tracks each opened file reader, instead of keeping a counter\footnote{\repolink{verifast/protocol-completion-1}}. As with this solution, when a file is initialized, it needs to be added to the list, and when a file is closed, it needs to be removed from the list. Additionally, the \textit{main} method requires a post-condition indicating that the ghost list is empty.

\subsubsection*{Sharing of mutable data}

Consider a scenario where some callbacks are executed asynchronously in a single-threaded context. This kind of use case is very common in web applications and \textit{Node.js}\footnote{\url{https://nodejs.org/en/}}, and works thanks to an \emph{event loop}, which is in charge of queuing and firing events without relying on multithreading (i.e. the event loop executes each callback at a time). Libraries such as \textit{ActiveJ}\footnote{\url{https://activej.io/}} may be used to implement asynchronous operations in Java.

Suppose that we have two callbacks that are responsible for adding a new item to a linked-list. Clearly both callbacks need exclusive access to the linked-list to modify it. Giving each callback access to the list at the point of initialization does not work because as soon as we split the permission, we only get read-only access. One could use locks to ensure mutual exclusion when modifying the list\footnote{\repolink{verifast/callbacks/CallbackWithLocks.java}}, but that would create unnecessary overhead since the lack of parallelism already ensures exclusive accesses.

The solution we found\footnote{\repolink{verifast/callbacks/EventLoop.java}}, suggested by Jacobs~\cite{private_conversation_bart_jacobs}, is to have a pool of objects that the event loop holds and provides to each callback at a time. For each callback to be able to find the linked-list in the pool, each holds 1/2 of the proof that the list is there. For simplicity, we implement an event loop that only allows two callbacks to be queued. For brevity, the listings containing Java code only show the methods' signature and contract.

For this implementation, we specify the \textit{eventloop} predicate (lines 3-7 of List.~\reflst{eventlooppredicates}) which holds access to the callbacks. Additionally, it keeps track of a ghost list containing predicate constructors and asserts that all the predicates hold, thanks to the combination of the \textit{foreach} (from VeriFast's library) and the \textit{holds} (line 1) predicates. We also define the \textit{listCtor} predicate constructor which, when applied, asserts that we hold full access to a given linked-list (line 9). The \textit{callback} predicate (lines 11-12) provides each callback access to its \textit{list} field and a fraction of the proof that the list's memory footprint is available in the ghost list.

\begin{minipage}{\linewidth}
\begin{lstlisting}[language=, caption=Event loop's predicates, label=\labellst{eventlooppredicates}]
predicate holds(predicate() p) = p();

predicate eventloop(EventLoop e, int id, list<predicate()> preds) =
  e.cb1 |-> ?cb1 &*& e.cb2 |-> ?cb2 &*&
  (cb1 == null ? emp : callback(cb1, id, _)) &*&
  (cb2 == null ? emp : callback(cb2, id, _)) &*&
  strong_ghost_list<predicate()>(id, preds) &*& foreach<predicate()>(preds, holds);

predicate_ctor listCtor(LinkedList l)() = l != null &*& llist(l, _, _, _);

predicate callback(Callback cb, int id, LinkedList l) =
  cb.list |-> l &*& [_]strong_ghost_list_member_handle(id, listCtor(l));
\end{lstlisting}
\end{minipage}

The \textit{EventLoop} class (List.~\reflst{eventloop}) provides 3 essential methods: \textit{addObject}, which inserts a new resource into the ghost list and provides a proof that such resource is in that list; \textit{addCallback}, which schedules a new callback to be called; and \textit{runAll}, which executes all queued callbacks, one at a time.

\begin{minipage}{\linewidth}
\begin{lstlisting}[language=Java, caption=Event loop's implementation, label=\labellst{eventloop}]
class EventLoop {  
  public EventLoop()
    //@ requires true;
    //@ ensures eventloop(this, _, nil);
  { ... }
  
  public void addObject(LinkedList l)
    //@ requires eventloop(this, ?id, ?preds) &*& listCtor(l)();
    /*@ ensures eventloop(this, id, append(preds, cons(listCtor(l), nil))) &*&
                strong_ghost_list_member_handle(id, listCtor(l)); */
  { ... }

  public void addCallback(Callback cb)
    //@ requires eventloop(this, ?id, ?preds) &*& callback(cb, id, _);
    //@ ensures eventloop(this, id, preds);
  { ... }
  
  public void runAll()
    //@ requires eventloop(this, ?id, ?preds);
    //@ ensures eventloop(this, id, preds);
  { ... }
}
\end{lstlisting}
\end{minipage}

When a callback is initialized, it requires a fraction of the proof that the linked-list's memory footprint is present in the event loop's ghost list (line 3 of List.~\reflst{callback}). This proof is kept in the \textit{callback}, as we have seen. When the callback is executed, with the \textit{run} method (line 7), the event loop provides its ghost list so that the callback can extract full access to the object it requires (line 8). After execution, the callback should return that access (line 9).

\begin{minipage}{\linewidth}
\begin{lstlisting}[language=Java, caption=Callback's implementation, label=\labellst{callback}]
class Callback {
  public Callback(LinkedList l)
    //@ requires [_]strong_ghost_list_member_handle(?id, listCtor(l));
    //@ ensures callback(this, id, l);
  { ... }

  public void run()
    //@ requires callback(this, ?id, ?l) &*& eventloop(?e, id, ?preds);
    //@ ensures callback(this, id, l) &*& eventloop(e, id, preds);
  { ... }
}
\end{lstlisting}
\end{minipage}

List.~\reflst{eventloopmain} presents a usage example of callbacks and the event loop. Initially, the shared linked-list and the event loop are initialized (lines 1-2). Then the memory footprint is provided to the event loop via the \textit{addObject} method (line 4). Following that, two callbacks are initialized and queued in the event loop (lines 7-10). Finally, both callbacks are executed (line 12).

\begin{minipage}{\linewidth}
\begin{lstlisting}[language=Java, caption=Callback's implementation, label=\labellst{eventloopmain}]
LinkedList l = new LinkedList();
EventLoop e = new EventLoop();
//@ close listCtor(l)();
e.addObject(l);

//@ split_fraction strong_ghost_list_member_handle(_, listCtor(l));
Callback3 c1 = new Callback3(l);
Callback3 c2 = new Callback3(l);
e.addCallback(c1);
e.addCallback(c2);

e.runAll();
\end{lstlisting}
\end{minipage}

Although we were able to successfully implement this use case, we believe this solution has some drawbacks. Firstly, we are forced to take all the resources the callbacks need (in this case, the linked-list) and put them explicitly in the pool of objects, which is cumbersome. Secondly, each callback needs to be aware of this pool and receive it in the pre-condition so that it can manipulate the required objects. In other words, instead of the event loop being abstracted away, its use is made explicit. We believe this highlights a very common scenario that occurs in tools that provide the possibility of deductive reasoning: even though it is usually possible to find a way to verify a complex application, the code needs to be implemented in such a way as to help the verifier check the specifications. This results in specifications that require reasoning that is not often related with the application's logic.

\subsubsection*{Assessment}

VeriFast's use of, namely, separation logic, fractional permissions, and predicates, allows for rich and expressive specifications that make it possible to verify complex programs. However, deductive reasoning is often required when the specifications are more elaborate. For example, the opening and closing of predicates is necessary regularly as well as the definition of multiple lemmas, as we have seen in the examples. This is tedious and can be a barrier to less experienced users.
In our experience, we spent more time in proving results than in writing the code, having had to write about 160 lines of lemmas to make the linked-list and iterator implementations work.\footnote{\repolink{verifast/basic/Lemmas.java}} However, implementing the file reader was very straightforward.

Although VeriFast has an IDE, which provides a way for one to look at each step of a proof when an error is discovered, we believe that, at least in part, the use of a proof assistance (similar to Coq\footnote{\url{https://coq.inria.fr/}} for example) would improve the user experience even more by avoiding having to insert proof guiding assertions in the code implementation itself (allowing for more separation of concerns) and avoiding the need to rerun the tool every time that occurs. In other words, when guiding the proofs, one would get immediate feedback. Nonetheless, the IDE was still very useful in helping us prove the aforementioned lemmas.

In the context of typestates, checking for protocol completion is crucial to ensure that necessary method calls are not forgotten and that resources are freed, thus avoiding memory leaks. Unfortunately, that concept is not built-in in separation logic. One solution we found is to have a counter that keeps track of all typestated-objects which are not in the final state. Whenever an object reaches the final state, the counter can be decremented. Then, we would have a post-condition in the {\it main} method saying the counter should be 0. Of course, this requires keeping hold of the aforementioned tracker in specifications, which can be a huge burden in bigger programs. Moreover, if one forgets to add the post-condition to the {\it main}, protocol completion will not be enforced. We believe that protocol completion should be provided directly by the type system and the programmer should not be required to remember to add this property to the specification. Ensuring protocol completion could be embedded in the logic and such a feature could even be possible in VeriFast. Given that VeriFast supports leak checking, one would just need to incorporate notions of typestates and ensure that leaking would only be allowed when objects are in their final states. In C, one would also need to enforce that claimed memory is freed. In Java, leak checking would need to be enabled for typestated-objects.

\newlabelscope

\subsection{VerCors}

\subsubsection*{File reader implementation}

To model the file reader's protocol, we use pre- and post-conditions in all public methods indicating the expected state and the destination state after the call. To track the current state we use an integer ghost field (line 2 of List.~\reflst{filereaderclass1}). The field {\it remaining} is the number of bytes left to read. When it is 0, it means we reached the end of the file.

\begin{minipage}{\linewidth}
\begin{lstlisting}[language=Java, caption={\it FileReader} class (part 1), label=\labellst{filereaderclass1}]
public class FileReader {
  //@ ghost private int state;
  private int remaining;
\end{lstlisting}
\end{minipage}

Consider, for example, the \textit{open} method (List.~\reflst{filereaderclass2}). The \textit{context} clause (line 1) specifies what is true in the pre-condition and post-condition. This is useful to avoid repetition. Therein we specify that we need full permission (i.e. fractional permission 1) to the \textit{state} and \textit{remaining} fields, and that the \textit{remaining} value is greater than or equal to zero. The separating conjunction binary operator is represented by the \texttt{**} symbol in VerCors. We could abstract this information in a predicate, but we keep it exposed to more easily track the \textit{state} and \textit{remaining} fields. Then we require that the object be in the \textit{Init} state, represented with number 1 (line 15). We tried to use constants, instead of explicitly using numbers, but there seems to be an internal error when static final fields are used in Java classes.\footnote{``[abort] (class vct.col.rewrite.Flatten)At file FileReader.java from line 2 column 23 until line 2 column 41: internal error: current block is null''} Then we ensure that the file changes to the \textit{Opened} state, represented with 2, and that the \textit{remaining} value is the same as before (line 17). The state change is performed with the ghost assignment in line 19. The state number for the \textit{Closed} state is 3. The rest of the implementation is very straightforward.\footnote{\repolink{vercors/FileReader.java}}

\begin{minipage}{\linewidth}
\begin{lstlisting}[language=Java, caption={\it FileReader} class (part 2), label=\labellst{filereaderclass2}, firstnumber=15]
  //@ context Perm(state, 1) ** Perm(remaining, 1) ** remaining >= 0;
  //@ requires state == 1;
  //@ ensures state == 2 ** remaining == \old(remaining);
  public boolean open() {
    //@ ghost this.state = 2;
    return true;
  }
  ...
}
\end{lstlisting}
\end{minipage}

\subsubsection*{Linked-list implementation}

To model the linked-list\footnote{\repolink{vercors/LinkedList.java}} we define in its class the {\it state} predicate (List.~\reflst{liststatepredicate}). This predicate holds the heap chunks of the {\it head} and {\it tail} fields and of all the nodes in the list.\footnote{They need to be called \textit{hd} and \textit{tl}, respectively, because \textit{head} and \textit{tail} are reserved words in VerCors.} The single parameter allows us to reason about the elements of the list in an abstract way. Lines 4 to 6 ensure that if one of the fields is \textit{null}, the other is also \textit{null} and the list is empty. To ease the addition of new elements to the list, which requires easy access to the tail node, we request access to the sequence of nodes between {\it head} (inclusive) and {\it tail} (exclusive), through the {\it nodes\_until} predicate (List.~\reflst{nodesuntilpredicate}), and then keep access to the {\it tail} directly in the body of the {\it state} predicate (line 8). Note that we do not hold permission to the fields of the values stored. This is to allow them to change independently of the linked-list.

\begin{minipage}{\linewidth}
\begin{lstlisting}[language=, caption={\it state} predicate of the linked-list, label=\labellst{liststatepredicate}]
final resource state(seq<FileReader> list) =
  Perm(hd, 1) ** Perm(tl, 1) **
  (
    hd == null ? (tl == null ** list == seq<FileReader> {}) :
      (
        tl == null ? (hd == null ** list == seq<FileReader> {}) :
          (
            nodes_until(hd, tl) ** PointsTo(tl.next, 1, null) ** Perm(tl.value, 1) **
            list == (nodes_until_list(hd, tl) + seq<FileReader> { tl.value })
          )
      )
  );
\end{lstlisting}
\end{minipage}

One may notice that we refer to the \textit{FileReader} type directly instead of making the \textit{LinkedList} generic over types of elements. The reason for this is that VerCors does not support generics in Java.

The {\it nodes\_until} predicate (List.~\reflst{nodesuntilpredicate}) holds the heap chunks of the nodes in the sequence starting on {\it n} (inclusive) and ending on {\it end} (exclusive). The {\it nodes\_until\_list} method is used to reason about the values that are stored in this sequence of nodes (List.~\reflst{nodesuntillistmethod}). Notice how {\it nodes\_until\_list} requires access to the heap chunk of nodes in the sequence from {\it n} to {\it end} (line 1 of List.~\reflst{nodesuntillistmethod}), and unfolds it to expose the permissions (line 3). This is necessary so that it can compute the corresponding sequence.

\begin{minipage}{\linewidth}
\begin{lstlisting}[language=, caption={\it nodes\_until} predicate, label=\labellst{nodesuntilpredicate}]
static resource nodes_until(Node n, Node end) =
  n == end ?
    true :
    n != null ** Perm(n.next, 1) ** Perm(n.value, 1) ** nodes_until(n.next, end);
\end{lstlisting}
\end{minipage}

\begin{minipage}{\linewidth}
\begin{lstlisting}[language=, caption={\it nodes\_until\_list} method, label=\labellst{nodesuntillistmethod}]
requires [1\2]nodes_until(n, end);
static pure seq<FileReader> nodes_until_list(Node n, Node end) =
  \unfolding [1\2]nodes_until(n, end) \in (
    n == end ?
      seq<FileReader> {} :
      seq<FileReader> { n.value } + nodes_until_list(n.next, end)
  );
\end{lstlisting}
\end{minipage}

The implementation of the {\it remove}\footnote{\repolink{vercors/LinkedList.java\#L183-L207}} and {\it isEmpty}\footnote{\repolink{vercors/LinkedList.java\#L209-L217}} methods is straightforward requiring only the unfolding and folding of the aforementioned predicates a few times. The implementation of the {\it add} method\footnote{\repolink{vercors/LinkedList.java\#L158-L181}} requires a lemma (List.~\reflst{addlemma}) stating that if we have a sequence of nodes plus the final node, and we append another node in the end, we get a new sequence with all the nodes from the previous sequence, the previous final node, and the newly appended node.

\begin{minipage}{\linewidth}
\begin{lstlisting}[language=, caption={\it add\_lemma} lemma, label=\labellst{addlemma}]
requires n1 != null ** n2 != null ** n3 != null;
requires nodes_until(n1, n2) ** Perm(n2.next, 1) ** Perm(n2.value, 1);
requires n2.next == n3 ** PointsTo(n3.next, 1, null);
requires list == nodes_until_list(n1, n2) + seq<FileReader> { n2.value };
ensures nodes_until(n1, n3) ** PointsTo(n3.next, 1, null) ** list == nodes_until_list(n1, n3);
ghost static void add_lemma(Node n1, Node n2, Node n3, seq<FileReader> list) {
  ...
}
\end{lstlisting}
\end{minipage}

The \textit{add} method (List.~\reflst{addmethod}) accepts in the real code a file reader (line 5). In the ghost code, it receives an additional parameter which is the current sequence of values in the list, with the \textit{given} clause (line 1). The real code returns nothing. The ghost code returns the new sequence of values, with the \textit{yields} clause (line 2). The method's contract requires that we have permission for the state of the linked-list, and that the sequence of values corresponds to the sequence which was given (line 3). Then the method ensures that the new sequence is built from the old one by appending the new added value (line 4). The use of the \textit{given} and \textit{yields} is essentially for users of this method to reason about the values in the list.

\begin{minipage}{\linewidth}
\begin{lstlisting}[language=Java, caption={\it add} method, label=\labellst{addmethod}]
//@ given seq<FileReader> oldList;
//@ yields seq<FileReader> newList;
//@ requires state(oldList);
//@ ensures state(newList) ** newList == oldList + seq<FileReader> { x };
public void add(FileReader x) {
  ...
}
\end{lstlisting}
\end{minipage}

Instead of using the \textit{given} and \textit{yields} clauses, we would have preferred to rely on a ghost field storing a sequence of file readers, to reason about those values in an abstract way. Unfortunately, it seems one cannot reason about the old value of a field if permission for that field is inside a predicate. Using inline unfolding does not seem to work (List.~\reflst{oldunfolding}).\footnote{\repolink{vercors/unfold-old}}

\begin{minipage}{\linewidth}
\begin{lstlisting}[language=, caption=Reasoning about old fields, label=\labellst{oldunfolding}]
  NotWellFormed:InsufficientPermission
=========================================
=== LinkedList.java               ===
 159   //@ requires state() ** \unfolding state() \in true;
                                                                 [----
 160   //@ ensures state() ** \unfolding state() \in list == \old(list) + seq<FileReader> { x };
                                                                  ----]
 161   public void add(FileReader x) {
\end{lstlisting}
\end{minipage}

\subsubsection*{Iterator implementation}

To model the iterator\footnote{\repolink{vercors/LinkedListIterator.java}} we define in its class the {\it state} predicate (List.~\reflst{iteratorstatepredicate}) which holds access to the current node in the {\it curr} field and all the nodes in the linked-list. The two final parameters are the list of elements already read and the list of elements still to read, respectively.
In this predicate, the permissions to the nodes are split in two parts. Half of the permissions ``preserves the structure'' of the list (line 4), and the other half holds the view of the iterator: a sequence of nodes from the {\it head} (inclusive) to the current node (exclusive), using the {\it nodes\_until} predicate (line 5); and a sequence from the current node (inclusive) to the final one, using the {\it nodes} predicate (line 6) (List.~\reflst{nodespredicate}).
For practical reasons, the {\it head} and {\it tail} are also referenced from the {\it first} and {\it last} ghost fields of the iterator (line 10).

\begin{minipage}{\linewidth}
\begin{lstlisting}[language=, caption={\it state} predicate of the iterator, label=\labellst{iteratorstatepredicate}]
final resource state(LinkedList linkedlist, seq<FileReader> s, seq<FileReader> r) =
  Perm(first, 1) ** Perm(last, 1) ** Perm(curr, 1) **
  Perm(linkedlist.hd, 1\2) ** Perm(linkedlist.tl, 1\2) **
  ([1\2]linkedlist.state(s + r)) **
  ([1\2]LinkedList.nodes_until(first, curr)) **
  ([1\2]LinkedList.nodes(curr)) **
  s == LinkedList.nodes_until_list(first, curr) **
  r == LinkedList.nodes_list(curr) **
  (\unfolding [1\2]linkedlist.state(s + r) \in
    (first == linkedlist.hd ** last == linkedlist.tl));
\end{lstlisting}
\end{minipage}

\begin{minipage}{\linewidth}
\begin{lstlisting}[language=, caption={\it nodes} predicate, label=\labellst{nodespredicate}]
static resource nodes(Node n) =
  n == null ? true : Perm(n.next, 1) ** Perm(n.value, 1) ** nodes(n.next);
\end{lstlisting}
\end{minipage}

To create the iterator\footnote{\repolink{vercors/LinkedList.java\#L219-L232}}, we take the permission to all the nodes in the linked-list and split it in the aforementioned way. After iterating through all the nodes, the full permission to the nodes needs to be restored to the list. These actions are done with the {\it prepare\_iterator} (List.~\reflst{prepareiteratorlemma}) and the {\it dispose\_iterator} (List.~\reflst{disposeiteratorlemma}) lemmas, respectively.

\begin{minipage}{\linewidth}
\begin{lstlisting}[language=, caption={\it prepare\_iterator} lemma, label=\labellst{prepareiteratorlemma}]
requires ([1\2]nodes_until(hd, tl)) ** PointsTo(tl.next, 1\2, null) ** Perm(tl.value, 1\2);
requires list == nodes_until_list(hd, tl) + seq<FileReader> { tl.value };
ensures [1\2]nodes(hd);
ensures list == nodes_list(hd);
ghost static void prepare_iterator(Node hd, Node tl, seq<FileReader> list) {
  ...
}
\end{lstlisting}
\end{minipage}

\begin{minipage}{\linewidth}
\begin{lstlisting}[language=, caption={\it dispose\_iterator} lemma, label=\labellst{disposeiteratorlemma}]
requires ([1\2]nodes_until(hd, tl)) ** PointsTo(tl.next, 1\2, null) ** Perm(tl.value, 1\2);
requires list == nodes_until_list(hd, tl) + seq<FileReader> { tl.value };
requires [1\2]nodes_until(hd, null);
requires list == nodes_until_list(hd, null);
ensures nodes_until(hd, tl) ** PointsTo(tl.next, 1, null) ** Perm(tl.value, 1);
ensures list == nodes_until_list(hd, tl) + seq<FileReader> { tl.value };
ghost static void dispose_iterator(Node hd, Node tl, seq<FileReader> list) {
  ...
}
\end{lstlisting}
\end{minipage}

The implementation of the {\it hasNext} method\footnote{\repolink{vercors/LinkedListIterator.java\#L52-L62}} is straightforward and requires only the unfolding and folding of the {\it state} predicate. The implementation of the {\it next} method\footnote{\repolink{vercors/LinkedListIterator.java\#L64-L82}} requires unfolding and folding predicates, and the {\it advance} lemma, which helps us advance the state of the iterator, moving the just retrieved value from the ``to see'' list to the ``seen'' list\footnote{\repolink{vercors/LinkedList.java\#L103-L114}}.

\subsubsection*{Protocol completion}

In an attempt to ensure that all file readers created through the lifetime of the program reach the end of their protocol, we define a \textit{FileTracker}\footnote{\repolink{vercors/FileTracker.java}} which keeps hold of the number of open file readers using ghost code. Then, we augment the file reader's implementation to increment this counter in the constructor (List.~\reflst{filereaderconstructor2}), and decrement the counter in the \textit{close} method (List.~\reflst{filereaderclose2}). Notice how the tracker is given to the constructor and to the \textit{close} method as a ghost parameter through the {\it given} directive (line 1 of both listings). Finally, we assert in the pre-condition and post-condition of the \textit{main} that the counter should be 0 (List.~\reflst{trackermain}).

\begin{minipage}{\linewidth}
\begin{lstlisting}[language=Java, caption=FileReader's constructor, label=\labellst{filereaderconstructor2}]
//@ given FileTracker tracker;
//@ context Perm(tracker.active, 1);
//@ ensures Perm(state, 1) ** Perm(remaining, 1) ** state == 1 ** remaining >= 0;
//@ ensures tracker.active == \old(tracker.active) + 1;
public FileReader() {
  //@ ghost this.state = 1;
  this.remaining = 20;
  //@ ghost tracker.inc();
}
\end{lstlisting}
\end{minipage}

\begin{minipage}{\linewidth}
\begin{lstlisting}[language=Java, caption=FileReader's \textit{close} method, label=\labellst{filereaderclose2}]
//@ given FileTracker tracker;
//@ context Perm(tracker.active, 1);
//@ context Perm(state, 1) ** Perm(remaining, 1) ** remaining >= 0;
//@ requires state == 2 ** remaining == 0;
//@ ensures state == 3;
//@ ensures tracker.active == \old(tracker.active) - 1;
public void close() {
  //@ ghost this.state = 3;
  //@ ghost tracker.dec();
}
\end{lstlisting}
\end{minipage}

\begin{minipage}{\linewidth}
\begin{lstlisting}[language=Java, caption={\it main} method, label=\labellst{trackermain}]
//@ given FileTracker tracker;
//@ context Perm(tracker.active, 1);
//@ requires tracker.active == 0;
//@ ensures tracker.active == 0;
public static void main(String[] args) {
  ...
}
\end{lstlisting}
\end{minipage}

Unfortunately, it is possible to fail to ensure protocol completion. Firstly, one could forget to increment and decrement the counter when the typestated-object is initialized and when its protocol finishes, respectively. Secondly, one could forget to add the post-condition to the {\it main} method. So, as with VeriFast (and in general, with deductive verification), we can guarantee protocol completion but only if the programmer does not fall for these ``traps''. Again we see that ghost code is useful, but if such code is not correctly connected with the ``real'' code, we actually guarantee nothing.

Since VerCors supports quantifiers, one could think of quantifying over all file readers and ensuring they are all closed. Unfortunately, we would be quantifying over all possible file readers, not just the ones actually allocated on the heap.

\subsubsection*{Use example}

To exemplify the use of a linked-list with file readers, we instantiate a linked-list and add three file readers in the initial state to it (lines 10-12 of List.~\reflst{main}). Then we pass the list to the {\it useFiles} method (line 16) which iterates through all the files and executes their protocol to the end. The ghost sequence \textit{l} will allow us to track the values in the list (line 1). The \textit{then} clauses receive the returned ghost list and assign it to the local variable \textit{l}. The \textit{with} clauses provide the needed ghost parameters to the respective calls. In this example, we created a \textit{filereader} predicate to just track the states and avoid verbosity (List.~\reflst{filereaderpredicate}). One may notice that we assert that all file readers are different (line 6). This is because \textit{forall} quantifiers, like the one in line 15, which talk about permissions need to be injective.\footnote{\url{https://vercors.ewi.utwente.nl/wiki/\#injectivity-object-arrays-and-efficient-verification}}

\begin{minipage}{\linewidth}
\begin{lstlisting}[language=Java, caption={\it Main} code, label=\labellst{main}]
//@ ghost seq<FileReader> l;
LinkedList list = new LinkedList() /*@ then {l=newList;} @*/;
FileReader f1 = new FileReader() /*@ with {tracker=tracker;} @*/;
FileReader f2 = new FileReader() /*@ with {tracker=tracker;} @*/;
FileReader f3 = new FileReader() /*@ with {tracker=tracker;} @*/;
//@ assert f1 != f2 && f2 != f3 && f1 != f3;
//@ fold filereader(f1, 1);
//@ fold filereader(f2, 1);
//@ fold filereader(f3, 1);
list.add(f1) /*@ with {oldList=l;} then {l=newList;} @*/;
list.add(f2) /*@ with {oldList=l;} then {l=newList;} @*/;
list.add(f3) /*@ with {oldList=l;} then {l=newList;} @*/;
//@ assert l == seq<FileReader> {f1, f2, f3};
//@ assert |l| == 3;
//@ assert (\forall* int i; i >= 0 && i < |l|; filereader(l[i], 1));
useFiles(list) /*@ with {tracker=tracker; l=l;} @*/;
\end{lstlisting}
\end{minipage}

\begin{minipage}{\linewidth}
\begin{lstlisting}[language=, caption={\it filereader} predicate, label=\labellst{filereaderpredicate}]
static resource filereader(FileReader f, int state) =
  PointsTo(f.state, 1, state) ** Perm(f.remaining, 1) ** f.remaining >= 0;
\end{lstlisting}
\end{minipage}

The pre- and post-conditions of the \textit{useFiles} method ensure that we get a list with all file readers in the initial state (line 7 of List.~\reflst{usefilescontract}), and we end up with the same list but with the readers in the final state (line 8). This is possible because the predicate which models the iterator keeps track of the elements already seen and the elements to see. The \textit{context\_everywhere} clause asserts that something holds in the pre-condition, post-condition, and in all the loop invariants inside the method (line 3).

\begin{minipage}{\linewidth}
\begin{lstlisting}[language=, caption={\it useFiles} contract, label=\labellst{usefilescontract}]
//@ given FileTracker tracker;
//@ given seq<FileReader> l;
//@ context_everywhere Perm(tracker.active, 1);
//@ requires tracker.active == |l|;
//@ ensures tracker.active == 0;
//@ context linkedlist != null ** linkedlist.state(l);
//@ requires (\forall* int i; i >= 0 && i < |l|; filereader(l[i], 1));
//@ ensures (\forall* int i; i >= 0 && i < |l|; filereader(l[i], 3));
public static void useFiles(LinkedList linkedlist)
\end{lstlisting}
\end{minipage}

The verification of the {\it useFiles} method (List.~\reflst{usefilescode}) makes use of two local ghost sequences to track the values seen and the remaining values to iterate through (lines 1-2). We also need to provide the appropriate ghost parameters and retrieve the ghost results in all the calls that make use of them. Furthermore, loop invariants need to be defined (lines 6-10 and lines 18-19), the unfolding (line 15) of the {\it filereader} predicate needs to be performed to get access to the \textit{remaining} field, and the folding (line 25) to restore the outer loop's invariant. Finally, we restore full permission from the iterator to the linked-list by calling the \textit{dispose} method on the iterator (line 28), which uses the {\it dispose\_iterator} lemma (List.~\reflst{disposeiteratorlemma}). The boolean results of the \textit{hasNext} and \textit{eof} calls are stored in temporary local variables, instead of being used directly in loops' conditions, to workaround an issue.\footnote{\url{https://github.com/utwente-fmt/vercors/issues/436}}

\begin{minipage}{\linewidth}
\begin{lstlisting}[language=Java, caption={\it useFiles} code, label=\labellst{usefilescode}]
//@ ghost seq<FileReader> seen = seq<FileReader> {};
//@ ghost seq<FileReader> remaining = l;
LinkedListIterator it = linkedlist.iterator() /*@ with {list=l;} @*/;
//@ assert it.state(linkedlist, seen, remaining);
boolean has = it.hasNext() /*@ with {linkedlist=linkedlist; s=seen; r=remaining;} @*/;
//@ loop_invariant it != null ** it.state(linkedlist, seen, remaining) ** seen + remaining == l;
//@ loop_invariant tracker.active == |remaining|;
//@ loop_invariant has == (|remaining| > 0);
//@ loop_invariant (\forall* int i; i >= 0 && i < |seen|; filereader(seen[i], 3));
//@ loop_invariant (\forall* int i; i >= 0 && i < |remaining|; filereader(remaining[i], 1));
while (has) {
  FileReader f = it.next() /*@ with {linkedlist=linkedlist; s=seen; r=remaining;} @*/;
  //@ ghost seen = seen + seq<FileReader> { f };
  //@ ghost remaining = tail(remaining);
  //@ unfold filereader(f, 1);
  f.open();
  boolean end = f.eof();
  //@ loop_invariant Perm(f.state, 1) ** Perm(f.remaining, 1) ** f.state == 2 ** f.remaining >= 0 ** (end == (f.remaining == 0));
  //@ loop_invariant tracker.active == |remaining| + 1;
  while (!end) {
    f.read();
    end = f.eof();
  }
  f.close() /*@ with {tracker=tracker;} @*/;
  //@ fold filereader(f, 3);
  has = it.hasNext() /*@ with {linkedlist=linkedlist; s=seen; r=remaining;} @*/;
}
it.dispose() /*@ with {linkedlist=linkedlist; list=seen;} @*/;
\end{lstlisting}
\end{minipage}

\subsubsection*{Sharing of mutable data}

Consider again the scenario where some callbacks are executed asynchronously in a single-threaded context. As we have seen for VeriFast\footnote{\repolink{verifast/callbacks/EventLoop.java}}, both callbacks need exclusive access to the linked-list to modify it. Giving each callback access to the list at the point of initialization does not work because as soon as we split the permission, we only get read-only access. We also want to avoid the unnecessary overhead of locks.

The solution we found for VeriFast was to have a pool of objects that the event loop holds and provides to each callback at a time. Then, when each callback is called, it would extract the list from the pool, use it, and then return it to the pool. To allow the pool to be generic over different types of objects, each callback would hold 1/2 of the proof that the list is in the pool. Unfortunately, as far as we can tell, there is no support for higher-order predicates, which we would need to store the permissions to the objects in the pool.

\subsubsection*{Assessment}

The use of separation logic, fractional permissions, and predicates, as in VeriFast, allows for rich and expressive specifications, but deductive reasoning is (again) required, by the definition of auxiliary lemmas, and the regular unfolding and folding of predicates. As we noted before, this can be tedious, as highlighted by the time and lines of code we needed to prove results, having had to write about 100 lines of lemmas.\footnote{\repolink{vercors/LinkedList.java\#L31-L142}}
Support for interactive proofs would be really helpful here.

In terms of user experience, we think allowing for more separation between lemma and predicate functions from code, instead of forcing these to belong to classes as static methods, would help improve readability, as others have also noted~\cite{Hollander2021}. We also believe that error messages could be more helpful\footnote{\url{https://github.com/utwente-fmt/vercors/issues/479}}, and the tool could be faster. Since the specifications are self-framing (i.e. only depend on memory locations that they themselves require to be accessible) and the checking process is modular, VerCors could make use of caching to avoid re-checking parts of the code that were not modified. We also noticed that if we unfolded a predicate on which some truth depends on, that knowledge would be lost. For example, the knowledge of the values stored in a sequence of nodes depends on the permissions to those nodes, available in the \codelink{vercors/LinkedList.java\#L14-L17}{\textit{nodes\_until} predicate}. In principle, unfolding this predicate should not invalidate the available information, but it does. To workaround this, we had to use fractional permissions to keep hold of some fraction of the original predicate, and only unfold the other fractional part.\footnote{\repolink{vercors/LinkedList.java\#L132}} Additionally, the need for unfolding predicates inline seems cumbersome.

\newlabelscope

\subsection{Comparison between VeriFast and VerCors}

Given the similarities between VeriFast and VerCors, we believe it is very relevant to provide a comparison between both. Other works also provide comparisons: \citet{DBLP:conf/icse-formalise/LathouwersH22} examine the annotation effort in several tools, including VeriFast and VerCors, and \citet{Hollander2021} briefly discusses both.

With respect to specifying access to memory locations, VeriFast only supports the \textbf{points-to assertions} of separation logic, while VerCors also supports \textbf{permission annotations}, following the approach of Chalice~\cite{leino2009basis,leino2009verification}, allowing us to refer to values in variables without the need to use new names for them. Furthermore, VerCors has built-in support for quantifiers, many different abstract data structures, and ghost code, which VeriFast does not. Nonetheless, VeriFast supports the definition of new inductive data types, fixpoint functions, higher-order predicates, and counting permissions, which VerCors does not. Unfortunately, VerCors does not support generics in Java.

\begin{minipage}{\linewidth}
\begin{lstlisting}[language=, caption={Inductive type and fixpoint example in VeriFast\protect\footnotemark}, label=\labellst{verifastexample}]
inductive list<t> = nil | cons(t, list<t>);

fixpoint t head<t>(list<t> xs) {
    switch (xs) {
        case nil: return default_value<t>;
        case cons(x, xs0): return x;
    }
}
\end{lstlisting}
\end{minipage}

\footnotetext{Example from \url{https://github.com/verifast/verifast/blob/master/bin/rt/\_list.javaspec}}

Both provide support for fractional permissions. However, this model only allows for read-only access when data is shared. In consequence, either locks are required to mutate shared data (even in single-threaded code, where they are not really necessary, resulting in inefficient code), or a complex specification workaround is needed. We believe that the specifications and code should focus on the application's logic, and the need to modify them to help the verifier should be avoided as much as possible. VerCors lacks support for counting permissions, which would allow permissions to be split in other ways.

When considering the deductive reasoning often required, we noted that more interactive proofs would improve the user experience. VeriFast already provides a way for one to observe each step of a proof when an error is encountered, and we believe VerCors would benefit from something similar, to avoid the need to practice ``trial and error'' constantly. In both tools, we spent more time in proving results about the linked-list and iterator than in writing the code. Some of the time spent in VerCors with the proofs was reduced because we could reuse the experience we had with VeriFast.

Finally, we missed the support for output parameters which VeriFast has. To reproduce the same concept in VerCors, we had to add ghost parameters in many methods and explicitly pass values for those parameters when calling such methods. For example, when working with the linked-list, we kept track of the sequence of values in the list through ghost code, and always had to pass that sequence to each called method.\footnote{\repolink{vercors/Main.java\#L16-L18}}

\begin{figure}
  \centering
  \includegraphics[width=0.5\linewidth]{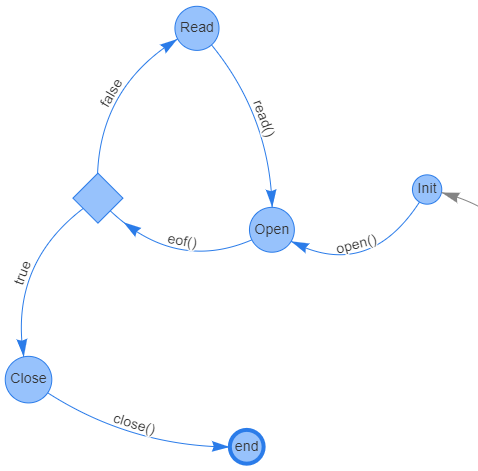}
  \caption{File reader's protocol}
  \label{fig:filereaderprotocol2}
\end{figure}

\newlabelscope

\subsection{Plural}

\subsubsection*{File reader implementation}

To model the file reader's protocol (Figure~\ref{fig:filereaderprotocol2}) in Plural, we define three states, {\it init}, {\it opened}, and {\it closed}, which refine the root state {\it alive} (line 4 of List.~\reflst{filereaderclass1}), a state that all objects have. Refining means we declare a state that is substate of the state we refine. Additionally, we define two states, {\it eof} and {\it notEof}, which refine the {\it opened} state, indicating if we have reached or not the end of the file, respectively (line 5). The file reader has a boolean field called {\it remaining}, indicating if there is still something to read. If we have reached the end of the file, {\it remaining} is {\it false}, otherwise it is {\it true}. The invariants associated with the states {\it eof} and {\it notEof} ensure these properties (lines 10-11). Ideally, the {\it remaining} field would contain an integer value (with the number of bytes to read), unfortunately the syntax {\it remaining == 0} does not seem to be supported in the invariants.

\begin{minipage}{\linewidth}
\begin{lstlisting}[language=, caption={\it FileReader} class (part 1), label=\labellst{filereaderclass1}]
import edu.cmu.cs.plural.annot.*;

@Refine({
  @States(value={"init", "opened", "closed"}, refined="alive"),
  @States(value={"eof", "notEof"}, refined="opened")
})
@ClassStates({
  @State(name="init", inv="opened == false * closed == false"),
  @State(name="opened", inv="opened == true * closed == false"),
  @State(name="eof", inv="remaining == false"),
  @State(name="notEof", inv="remaining == true"),
  @State(name="closed", inv="opened == true * closed == true")
})
public class FileReader {
\end{lstlisting}
\end{minipage}

The \textit{open}, \textit{read}, and \textit{close} methods require \textit{unique} permission to the receiver object to change the state. \textit{Full} permissions would be enough except for the possibility of concurrent accesses, which would require the methods to be {\it synchronized}.\footnote{We could disable the synchronization checker, but we do not think that is the correct and safe approach.}
The \textit{use=Use.FIELDS} clause indicates that the methods access fields but do not perform dynamically dispatched calls.
The \textit{eof} method only needs an \textit{immutable} permission guaranteeing that we are in the \textit{opened} state, and returns a boolean value indicating if the end of the file was reached (lines 36-44). The {\it if} statement (line 40) is required for Plural to understand that if {\it remaining} is {\it false}, we are in fact in the {\it eof} state. This implementation was accepted by Plural.\footnote{\repolink{plural/FileReader.java}}

\begin{minipage}{\linewidth}
\begin{lstlisting}[language=, caption={\it FileReader} class (part 2), label=\labellst{filereaderclass2}, firstnumber=25]
  @Unique(requires="init", ensures="opened", use=Use.FIELDS)
  public void open() {
    opened = true;
  }

  @Unique(requires="notEof", ensures="opened", use=Use.FIELDS)
  public byte read() {
    remaining = false;
    return 0;
  }

  @Imm(guarantee="opened", use=Use.FIELDS)
  @TrueIndicates("eof")
  @FalseIndicates("notEof")
  public boolean eof() {
    if (remaining == false)
      return true;
    return false;
  }
\end{lstlisting}
\end{minipage}

\subsubsection*{File reader examples}

In List.~\reflst{pluralfilereaderexample1}, we have an example of correct use of the file reader. In this scenario, Plural reports no errors, as expected.

\begin{minipage}{\linewidth}
\begin{lstlisting}[language=Java, caption=Correct use of file reader, label=\labellst{pluralfilereaderexample1}]
FileReader f = new FileReader();
f.open();
while (!f.eof()) f.read();
f.close();
\end{lstlisting}
\end{minipage}

In List.~\reflst{pluralfilereaderexample1}, we show an example of incorrect use of the file reader. Notice how the loop condition (line 3) is inverted and we are calling {\it read} when we reach the end of the file, instead of doing the opposite. Additionally, this causes {\it close} to be called before reaching the end of the file. Plural, as expected, reports these errors in lines 4 and 6.

\begin{minipage}{\linewidth}
\begin{lstlisting}[language=Java, caption=Incorrect use of file reader, label=\labellst{pluralfilereaderexample2}]
FileReader f = new FileReader();
f.open();
while (f.eof()) {
  f.read(); // error: argument this must be in state [notEof] but is in [eof]
}
f.close(); // error: argument this must be in state [eof] but is in [notEof]
\end{lstlisting}
\end{minipage}

Finally, in List.~\reflst{pluralfilereaderexample3}, we have a method that expects to receive a file reader in the {\it opened} state and does not return it to the caller (notice the {\it returned} parameter in the annotation). In this scenario, Plural reports no errors even though we are ``dropping'' a permission to the file reader. As far as we can tell, Plural's specification language is based on linear logic, which would imply that this scenario would not be accepted. However, we understand why this is the case in Java, since it is common for Java programmers to stop using an object and letting the garbage collector reclaim memory. Nonetheless, we believe that ensuring protocol completion is crucial for typestated-objects, to ensure that necessary method calls are not forgotten and resources are freed (e.g. closing a socket). Unfortunately, protocol completion is not a guarantee that Plural seems to provide.

\begin{minipage}{\linewidth}
\begin{lstlisting}[language=Java, caption=Dropping file reader, label=\labellst{pluralfilereaderexample3}]
void droppingObject(
  @Unique(requires="opened", returned=false) FileReader f) {}
\end{lstlisting}
\end{minipage}

\subsubsection*{Linked-list implementation}

The linked-list requires a \textit{Node}\footnote{\repolink{plural/Node.java}} and \textit{LinkedList}\footnote{\repolink{plural/LinkedList.java}} classes. These were adapted from a stack example provided in Plural's repository.\footnote{File pluralism/trunk/PluralTestsAndExamples/src/edu/cmu/cs/plural/polymorphism/ecoop/Stack.java} Naturally, we made the appropriate changes since our linked-list follows a FIFO discipline, while a stack follows a LIFO one.

To model each node we define two orthogonal state dimensions, {\it dimValue} and {\it dimNext}, which handle the {\it value} and {\it next} fields, respectively (lines 5 and 6 of List.~\reflst{plurallinkedlistnode}). In the {\it dimValue} dimension there are two states, {\it withValue} and {\it withoutValue}, which indicate if the node has permission to the stored value or not, respectively. In the {\it dimNext} dimension there are two states, {\it withNext} and {\it withoutNext}, which indicate if the node has permission to the next node or if {\it next} is {\it null}, respectively. The use of state dimensions was particularly useful to avoid the need to reason about all the combinations of having (or not) a value and having (or not) a next node. The invariants of each state are specified in lines 9 to 12.

The constructor accepts a (parametric) permission to a value and creates a node in states {\it withValue} and {\it withoutNext} (lines 19-23). The parametric permission \textit{p} is specified in the \textit{Similar} annotation in line 3.
The {\it getNext} method extracts the next node leaving the current one without a reference and permission to it (lines 25-32). The {\it setNext} method takes complete ownership of a node and sets it as the next node (lines 34-39). The {\it getValue} method gives all the permission it has to the value to the caller (lines 41-45). The {\it hasNext} method indicates if this node as a next node (lines 47-55). These methods are necessary because direct access to fields is only possible if the receiver object is \textit{this}.

%\begin{minipage}{\linewidth}
\begin{lstlisting}[language=Java, caption=Linked-list node implementation, label=\labellst{plurallinkedlistnode}]
import edu.cmu.cs.plural.annot.*;

@Similar("p")
@Refine({
  @States(value={"withValue", "withoutValue"}, dim="dimValue"),
  @States(value={"withNext", "withoutNext"}, dim="dimNext")
})
@ClassStates({
  @State(name="withValue", inv="p(value)"),
  @State(name="withoutValue", inv="true"),
  @State(name="withNext", inv="unique(next) in withValue,dimNext * next != null"),
  @State(name="withoutNext", inv="next == null")
})
public class Node<T> {
  @Apply("p")
  private Node<T> next;
  private T value;

  @Perm(ensures="unique(this!fr) in withValue,withoutNext")
  public Node(@PolyVar(value="p", returned=false) T val) {
    value = val;
    next = null;
  }

  @Unique(requires="withNext", ensures="withoutNext", use=Use.FIELDS)
  @ResultUnique(ensures="withValue,dimNext")
  @ResultApply("p2")
  public Node<T> getNext() {
    Node<T> n = next;
    next = null;
    return n;
  }

  @Unique(requires="withoutNext", ensures="withNext", use=Use.FIELDS)
  public void setNext(
    @Unique(requires="withValue,dimNext", returned=false) @Apply("p") Node<T> n
  ) {
    next = n;
  }

  @Unique(requires="withValue", ensures="withoutValue", use=Use.FIELDS)
  @ResultPolyVar("p")
  public T getValue() {
    return value;
  }

  @Unique(guarantee="dimNext", use=Use.FIELDS)
  @TrueIndicates("withNext")
  @FalseIndicates("withoutNext")
  public boolean hasNext() {
    if (next == null)
      return false;
    return true;
  }
}
\end{lstlisting}
%\end{minipage}

To model the list we define two states: the empty and the non-empty (line 5 of List.~\reflst{plurallinkedlist}). When it is empty, the {\it head} and {\it tail} fields are {\it null} (line 8). When it is not empty, {\it head} and {\it tail} are not {\it null} and there is unique permission to the first node, which is pointed by {\it head} (line 11). Since the {\it head} points to the next node, and so on, we should have the required chain of nodes that builds the linked-list.
To add a new element to the list, we require some (parametric) permission to the value to be added, then we create a new node, and append it to the end of the list (lines 26-36).
To remove an element from the list, we require the list to be non-empty, return the value stored in the \textit{head} node, and make the second node the new \textit{head} (lines 38-49).
Both operations require \textit{unique} permission to the list.

%\begin{minipage}{\linewidth}
\begin{lstlisting}[language=Java, caption=Linked-list implementation, label=\labellst{plurallinkedlist}]
import edu.cmu.cs.plural.annot.*;

@Similar("p")
@Refine({
  @States(value={"empty", "notEmpty"}, refined="alive")
})
@ClassStates({
  @State(name="empty", inv="head == null * tail == null"),
  @State(
    name="notEmpty",
    inv="unique(head) in withValue,dimNext * head != null * tail != null"
  )
})
public class LinkedList<T> {
  @Apply("p")
  private Node<T> head;
  @Apply("p")
  private Node<T> tail;

  @Perm(ensures="unique(this!fr) in empty")
  public LinkedList() {
    head = null;
    tail = null;
  }

  @Unique(requires="alive", ensures="notEmpty", use=Use.FIELDS)
  public void add(@PolyVar(value="p", returned=false) T value) {
    @Apply("p") Node<T> n = new Node<T>(value);
    if (head == null) {
      head = n;
      tail = n;
    } else {
      tail.setNext(n);
      tail = n;
    }
  }

  @Unique(requires="notEmpty", ensures="alive", use=Use.FIELDS)
  @ResultPolyVar("p")
  public T remove() {
    T result = head.getValue();
    if (head.hasNext()){
      head = head.getNext();
    } else {
      head = null;
      tail = null;
    }
    return result;
  }
}
\end{lstlisting}
%\end{minipage}

Unfortunately, Plural did not accept either implementation. With regards to the {\it Node} class, we had errors in all the methods indicating that the receiver could not be packed~\cite{deline2004typestates} to match the state specified by the {\it ensures} annotation parameter. Additionally, the invariant for the {\it withValue} state (i.e. {\it p(value)}, line 9 of List.~\reflst{plurallinkedlistnode}) had an error stating that the permission kind {\it p} was unknown, even though we introduced that parametric permission kind using the {\it @Similar} annotation (line 3). In fact, we did the same for the {\it LinkedList} class and we did not get these kinds of errors. We do not know the explanation for this issue.

With regards to the {\it LinkedList}, the only errors reported were in the {\it add} method. The {\it remove} method was accepted because we use {\it head.hasNext()} instead of {\it head != tail} to check if the list has more than one node, which directly informs Plural that there is another node next to the {\it head}.
To understand why {\it add} was not accepted, consider the case in which the list is not empty. In this scenario, the \textit{tail} is non-null, and we must call \textit{setNext} on it to append a new node to the list (line 33 of List.~\reflst{plurallinkedlist}). However, we need permission to do that and we do not have it.
For this example to work, we would need to obtain the permission to the last node that is owned by the second to last node. The VeriFast implementation of a linked-list\footnote{\repolink{verifast/basic/LinkedList.java}} required the definition of multiple predicates and lemmas. Since that does not seem to be supported by Plural, we think we cannot model the linked-list in this way.

An alternative solution could be to use \textit{share} permissions instead of \textit{unique} permissions in the nodes. Unfortunately, this would require locking when accessing them, because of the possibility of thread concurrency. Furthermore, we would lose track of the memory footprint used by the linked-list (since \textit{share} permissions allow for unrestricted aliasing). This can be an issue if we want to track all the references and ensure statically that all resources are freed at end.

Even if implementing such a structure was possible, we would like to have a collection that was parametric on the typestates of the objects stored inside, so that we could track the state changes of these. But, as far as we can tell, even though parametric fractional permissions are supported, the typestates need to be uniform.

\subsubsection*{Assessment}

Plural has the richest set of access permissions we know of, allowing the state of objects to be tracked even in the presence of aliasing, and permitting read/write and write/write operations, thanks to state guarantees. Nonetheless, we believe it is difficult to specify pointer-based data structures, such as linked-lists, where the last node is referenced from both the \textit{next} field of the second to last node and the \textit{tail} field.
We believe the support for logical predicates would be useful for specifying structures with recursive properties, and avoiding the repetition of statements. Furthermore, as far as we can tell, there is no support for parametric typestates, even though there is for fractional permissions, which would allow one to model a list with objects in different states that evolve.

The use of \textit{share} permissions allows for unrestricted aliasing. Nonetheless, state assumptions need to be discarded because of the possibility that there might be other threads attempting to modify the same reference. Although this thread-sharedness approximation is sound, it forces the use of synchronization primitives even if a reference is only available in one thread. \citet{DBLP:conf/oopsla/BeckmanBA08} discuss the possibility of distinguishing permissions for references that are only aliased locally from references that are shared between multiple threads, allowing access to thread-local ones without the need for synchronization. But as far as we know, the idea was not realized.

Furthermore, there seems to be no built-in support for guaranteeing protocol completion. This feature is important in typestate-oriented contexts because we want to ensure that necessary method calls are not forgotten and resources are freed. Nonetheless, this could be supported by asking the programmer to indicate which state should be the final state of a given object and allowing permissions (only) for \emph{ended} objects to be ``dropped''.

With respect to the programming effort, we do not think it was very demanding. We spent some time trying to understand which were the correct annotations to use (from examples in Plural's source code), and thinking about how to model the protocol of each class, but besides that the specification effort was minimal. However, this may be due to the fact that the file reader was very straightforward to implement, and because we knew beforehand that we would not be able to implement the linked-list.

\newlabelscope

\subsection{KeY}

\subsubsection*{File reader implementation}

To model the file reader's protocol\footnote{\repolink{key/FileReader.java}}, we use pre- and post-conditions in all public methods indicating the expected state and the destination state after the call. To track the current state we use an integer ghost field (line 6 of List.~\reflst{filereaderclass1}). We use constants to identify different states, thus avoiding the use of literal numbers in specifications (lines 2-4). The field {\it remaining} is the number of bytes left to read. When it is 0, it means we reached the end of the file. Its value should always be equal or greater than zero, as enforced by the invariant in line 14. The \textit{spec\_public} annotation allows these fields to be visible in specifications outside the \textit{FileReader} class even though they are private to code outside of the class.

We also declare the class as \textit{final} (line 1) to disallow the existence of subclasses. This will allow us to open the invariant in proofs when necessary. Such would not be possible in the presence of subclasses because the dynamic type of an object is not generally known statically and a unknown subclass could potentially add more invariants.

Since KeY uses first-order dynamic logic to verify programs, instead of separation logic for instance, the proofs explicitly mention the heap and how it is modified. Thus, each object should have a footprint, a set of locations that represent its state, and each of its methods should specify which locations may be modified and if the footprint itself is changed. This implementation technique is based on \textbf{dynamic frames}~\cite{DBLP:conf/fm/Kassios06}. In this example, we declare a set of locations and call it \textit{footprint} (line 9), and specify that it is composed by fields \textit{state} and \textit{remaining} (line 11). The \textit{footprint} is \textit{public} which means that its content can be opened during proofs. The \textit{accessible} annotation in line 10 specifies that \textit{footprint} (i.e. the set itself) will only change if a location in the set mentioned on the right-side of the colon changes. In this particular example, the \textit{footprint} is always the same, so we specify that nothing changes it.\footnote{\textit{\textbackslash nothing} denotes the empty set of locations.}

\begin{minipage}{\linewidth}
\begin{lstlisting}[language=Java, caption={\it FileReader} class (part 1), label=\labellst{filereaderclass1}]
public final class FileReader {
  /*@ ghost public static final int STATE_INIT = 1;*/
  /*@ ghost public static final int STATE_OPENED = 2;*/
  /*@ ghost public static final int STATE_CLOSED = 3;*/

  /*@ ghost private spec_public int state;*/
  private /*@ spec_public @*/ int remaining;

  /*@ public model \locset footprint;
    @ accessible footprint: \nothing;
    @ represents footprint = state, remaining;
    @*/
    
  /*@ invariant remaining >= 0; @*/
\end{lstlisting}
\end{minipage}

The constructor's specification indicates that it will not throw exceptions (\textit{normal\_behavior} annotation in line 17 of List.~\reflst{filereaderclass2}), it will not modify (i.e. assign to) any location that exists when it is called (line 18), it ensures that the initial state is \textit{Init} (line 19), and that the set of locations in footprint is now new, thus disjoint from other heap locations (line 20). The \textit{open} method requires that the file reader is in the \textit{Init} state (line 29), and ensures the new state is \textit{Opened} (line 31). The state change is performed with the \textit{set} ghost instruction (line 34). To modify the state, we also need to indicate that the footprint is assignable (line 30). Given that the file reader's footprint is very simple and declared with \textit{public} visibility, we may omit a post-condition indicating that the footprint does not change. The rest of the implementation is similar.\footnote{\repolink{key/FileReader.java}}

\begin{minipage}{\linewidth}
\begin{lstlisting}[language=Java, caption={\it FileReader} class (part 2), label=\labellst{filereaderclass2}, firstnumber=16]
  /*@
    @ normal_behavior
    @ assignable \nothing;
    @ ensures state == STATE_INIT;
    @ ensures \fresh(footprint);
    @*/
  public FileReader() {
    /*@ set state = STATE_INIT;*/
    this.remaining = 20;
  }

  /*@
    @ normal_behavior
    @ requires state == STATE_INIT;
    @ assignable footprint;
    @ ensures state == STATE_OPENED;
    @*/
  public void open() {
    /*@ set state = STATE_OPENED;*/
  }
  ...
}
\end{lstlisting}
\end{minipage}

To verify this class, we have to prove each method correct, according to each specification, and the fact that nothing changes the footprint (line 10). Each of these proofs is done separately since KeY provides modular verification: the facts in the pre-conditions should be enough to show the post-conditions hold. Given the simplicity of this class, it was not necessary to guide the proofs, so all the methods were automatically verified using KeY's default strategy. All the proof files showing the verification proof, for this example and the following ones, are available online.\footnote{\repolink{key/proofs}}

\subsubsection*{Single file reader usage}

Listing~\reflst{fileusage} presents a method\footnote{\repolink{key/Main.java\#L79-L98}} that takes a file reader in the \textit{Init} state and uses it until it reaches the \textit{Closed} state, as indicated by the pre- and post-conditions (lines 4 and 7). To verify the use of the file reader, it is necessary that its invariant holds. Such fact needs to be stated explicitly in the pre-condition using the \textit{\textbackslash invariant\_for} annotation (line 3). For client users of this code, we state that only the file reader's footprint is modified (line 5). For the loop to be verified, we need to provide some conditions. First, we specify that the file reader's invariant holds for the entirely of the loop's execution and that it is in the \textit{Opened} state (line 12). Then we indicate that only the file's footprint is modified (line 13). Finally, we provide \textit{remaining} as the decreasing value (line 14), thus ensuring termination.

\begin{minipage}{\linewidth}
\begin{lstlisting}[language=Java, caption=Single file reader usage, label=\labellst{fileusage}]
/*@
  @ normal_behavior
  @ requires \invariant_for(f);
  @ requires f.state == FileReader.STATE_INIT;
  @ assignable f.footprint;
  @ ensures \invariant_for(f);
  @ ensures f.state == FileReader.STATE_CLOSED;
  @*/
private static void use(FileReader f) {
  f.open();
  /*@
    @ loop_invariant \invariant_for(f) && f.state == FileReader.STATE_OPENED;
    @ assignable f.footprint;
    @ decreases f.remaining;
    @*/
  while (!f.eof()) {
    f.read();
  }
  f.close();
}
\end{lstlisting}
\end{minipage}

To verify this code, there was no need to guide the proof: KeY generated it fully automatically. Unfortunately, we only succeeded after many attempts at finding the correct specification. Recall that, as stated in the method's contract, only the file's footprint should be modified (line 5). But to prove that, we actually need to show that the footprint itself is not modified. This seems to be necessary because, in the \textbf{dynamic frames}~\cite{DBLP:conf/fm/Kassios06} approach, one may have heap-dependent expressions in assertions. One would expect that declaring that nothing changes the footprint would be enough (line 10 of List.~\reflst{filereaderclass1}), but sadly that does not seem to be the case. Before we found out that we could make the footprint public, we had to indicate in the post-conditions of file reader's methods that the footprint was not modified by them. By carrying this information with each method call, we could establish that the footprint itself remained unchanged. Now, with the footprint public, we can drop the extraneous post-conditions.

It is important to note that, in general, each method of an object should indicate if and how the footprint itself is modified. But in the file reader's case, since the footprint is a fixed set composed of two field locations, that is not necessary if we make it public. In any case, making the footprint public ended up being very useful in other examples to allow KeY to prove that, for example, the footprints of two objects of unrelated classes were disjoint. Although this seems an obvious fact, it was not always easy to show it, as we will demonstrate later.

\subsubsection*{Linked-list implementation}

The linked-list implementation\footnote{\repolink{key/LinkedList.java}} was heavily inspired in a tutorial by~\citet{DBLP:series/lncs/HiepBBG20}, except we did not implement a doubly-linked-list. To model it we declare several fields (List.~\reflst{linkedlistclass1}): \textit{size}, to count the number of values; \textit{head} and \textit{tail}, which point to the first and last nodes, respectively, or \textit{null}, in case the list is empty; \textit{nodeList}, containing a sequence of nodes; and \textit{values}, containing a sequence of values. The values the linked-list is going to store are file readers. Unfortunately, KeY does not support generics\footnote{\url{https://www.key-project.org/docs/user/RemoveGenerics/}} so we cannot take a parametric type instead. We could use the \textit{Object} type, which in fact would match Java's semantics (since generics are implemented by erasure\footnote{\url{https://docs.oracle.com/javase/tutorial/java/generics/erasure.html}}), but for the purposes of our example, we use the \textit{FileReader} type.

\begin{minipage}{\linewidth}
\begin{lstlisting}[language=Java, caption={\it LinkedList} class (part 1), label=\labellst{linkedlistclass1}]
public final class LinkedList {
  private /*@ spec_public @*/ int size;
  /*@ nullable @*/ private /*@ spec_public @*/ Node head;
  /*@ nullable @*/ private /*@ spec_public @*/ Node tail;
  //@ private spec_public ghost \seq nodeList;
  //@ private spec_public ghost \seq values;
\end{lstlisting}
\end{minipage}

As we did for the file reader, we define the linked-list's footprint (List.~\reflst{linkedlistclass2}). Such footprint is composed by all the list's fields (line 11) and the fields of each node in the sequence (line 12). We also specify that the footprint itself only changes if the sequence of nodes changes (line 9), and that the list's invariant may only change if the locations in the footprint change. Notice how the footprints of the values are not part of list's footprint. This is to allow the values to change independently of the linked-list. The proof that the footprint only depends on the nodes sequence was generated automatically by KeY using the default proof search strategy. Showing that the invariant only depends on the locations in the footprint required some interactivity to guide the proof.

\begin{minipage}{\linewidth}
\begin{lstlisting}[language=Java, caption={\it LinkedList} class (part 2), label=\labellst{linkedlistclass2}, firstnumber=8]
  /*@ public model \locset footprint;
    @ accessible footprint: nodeList;
    @ accessible \inv: footprint;
    @ represents footprint = size, head, tail, nodeList, values,
    @    (\infinite_union \bigint i; 0 <= i < nodeList.length; ((Node)nodeList[i]).*);
    @*/
\end{lstlisting}
\end{minipage}

The list's invariant is the most verbose part of the implementation (List.~\reflst{linkedlistclass3}). First, we specify that \textit{size} is equal to the number of nodes, which is then equal to the number of values (lines 16-17). We also specify that this number cannot overflow (line 18). Then we enforce that the values in the list are not \textit{null} (lines 19-22). One may notice the use of an existential quantifier, indicating that in each position of the sequences, elements exist. This is necessary because KeY treats sequences in a way where they may occasionally contain not-yet-created objects. Additionally, since the sequence declarations (lines 5-6 in List.~\reflst{linkedlistclass1}) do not enforce the type of their elements, we need to do it explicitly, either using the \textit{instanceof} operator, or using an existential quantifier. Furthermore, we need to cast the result of accessing a position in a given sequence. Following that, we have to take into account that the list may be empty. So, we define that either the nodes sequence is empty, and the \textit{head} and \textit{tail} fields are \textit{null} (line 23), or the nodes sequence is not empty, and the \textit{head} points to the first node, and \textit{tail} points to the last one (lines 24-26). To ensure we actually have a linked-list, we enforce that the \textit{next} field of each node points in fact to the following node in the sequence (lines 27-28). We also need to ensure that all the nodes are distinct (lines 29-32). Finally, we specify that each value in the values sequence corresponds to the value stored in each node in the same position (lines 33-34).

\begin{minipage}{\linewidth}
\begin{lstlisting}[language=Java, caption={\it LinkedList} class (part 3), label=\labellst{linkedlistclass3}, firstnumber=15]
  /*@ invariant
    @   nodeList.length == size &&
    @   nodeList.length == values.length &&
    @   nodeList.length <= Integer.MAX_VALUE &&
    @   (\forall \bigint i; 0 <= i < nodeList.length;
    @       (\exists Node n; n == nodeList[i] && n.value != null)) &&
    @   (\forall \bigint i; 0 <= i < values.length;
    @       (\exists FileReader f; f == values[i] && f != null)) &&
    @   ((nodeList == \seq_empty && head == null && tail == null)
    @    || (nodeList != \seq_empty && head != null && tail != null &&
    @         tail.next == null && head == (Node)nodeList[0] &&
    @         tail == (Node)nodeList[nodeList.length-1])) &&
    @   (\forall \bigint i; 0 <= i < nodeList.length-1;
    @       ((Node)nodeList[i]).next == (Node)nodeList[i+1]) &&
    @   (\forall \bigint i; 0 <= i < nodeList.length;
    @     (\forall \bigint j; 0 <= j < nodeList.length;
    @       (Node)nodeList[i] == (Node)nodeList[j] ==> i == j
    @   )) &&
    @   (\forall \bigint i; 0 <= i < values.length;
    @      values[i] == ((Node)nodeList[i]).value);
    @*/
\end{lstlisting}
\end{minipage}

The constructor's contract (List.~\reflst{linkedlistclass4}) specifies that no locations that are not fresh are modified (line 39), that the initial sequence of values is empty (line 40), and that the list's footprint is fresh (line 41). Verification was done automatically using the default strategy.

\begin{minipage}{\linewidth}
\begin{lstlisting}[language=Java, caption=Constructor's contract, label=\labellst{linkedlistclass4}, firstnumber=37]
  /*@
    @ normal_behavior
    @ assignable \nothing;
    @ ensures values == \seq_empty;
    @ ensures \fresh(footprint);
    @*/
\end{lstlisting}
\end{minipage}

The \textit{add} method\footnote{\repolink{key/LinkedList.java\#L51-L69}} stores a new value at the end of the list. Its contract (List.~\reflst{linkedlistclass5}) requires that there must be space to store one more element (line 53), and that only the list's footprint is assignable (line 54). It also indicates that, after the method call, there is a new value at the end of the sequence of values (line 55), and that any locations in the footprint either were present in the old footprint, or are fresh locations (line 56). The verification of the \textit{add} method was also done automatically using the default strategy.

\begin{minipage}{\linewidth}
\begin{lstlisting}[language=Java, caption={\it add} method's contract, label=\labellst{linkedlistclass5}, firstnumber=51]
  /*@
    @ normal_behavior
    @ requires values.length + (\bigint)1 <= Integer.MAX_VALUE;
    @ assignable footprint;
    @ ensures values == \seq_concat(\old(values), \seq_singleton(value));
    @ ensures \new_elems_fresh(footprint);
    @*/
\end{lstlisting}
\end{minipage}

The \textit{remove} method\footnote{\repolink{key/LinkedList.java\#L71-L92}} extracts a value from the start of the list. Its contract (List.~\reflst{linkedlistclass6}) requires that the list cannot be empty (line 73), and that only the list's footprint is assignable (line 74). After the call, it ensures the new sequence of values is the result of taking the old sequence and removing the first element (line 76), the value returned by the method is such element (line 77), and that the new footprint is a subset of the old one (line 78). The verification of the \textit{remove} method required some interactivity, namely to show that the first value was the value of the \textit{head}.

\begin{minipage}{\linewidth}
\begin{lstlisting}[language=Java, caption={\it remove} method's contract, label=\labellst{linkedlistclass6}, firstnumber=71]
  /*@
    @ normal_behavior
    @ requires values.length > 0;
    @ assignable footprint;
    @ ensures
    @   values == \seq_sub(\old(values), 1, \old(values).length) &&
    @   \result == (FileReader)\old(values)[0];
    @ ensures \subset(footprint, \old(footprint));
    @*/
\end{lstlisting}
\end{minipage}

The \textit{iterator} method\footnote{\repolink{key/LinkedList.java\#L102-L112}} returns a new iterator for this list (List.~\reflst{linkedlistclass7}). Its contract specifies that no locations that are not fresh are modified (line 96), that the iterator's invariant holds (line 97), that the iterator and its footprint are fresh (line 98), that the sequence of values iterated through is empty, that the sequence of values still to iterate is equal to the sequence of all values of the list (line 99), and that the iterator is associated with this list (line 100). The iterator's implementation will be presented in more detail in the following section. The verification of the \textit{iterator} method was performed automatically using the default strategy.

\begin{minipage}{\linewidth}
\begin{lstlisting}[language=Java, caption={\it iterator} method, label=\labellst{linkedlistclass7}, firstnumber=94]
/*@
  @ normal_behavior
  @ assignable \nothing;
  @ ensures \invariant_for(\result);
  @ ensures \fresh(\result) && \fresh(\result.footprint);
  @ ensures \result.seen == \seq_empty && \result.to_see == values;
  @ ensures \result.list == this;
  @*/
public /*@ pure @*/ LinkedListIterator iterator() {
  return new LinkedListIterator(this, head);
}
\end{lstlisting}
\end{minipage}

\subsubsection*{Iterator implementation}

To model the iterator\footnote{\repolink{key/LinkedListIterator.java}} we declare several fields (List.~\reflst{iteratorclass1}): \textit{list}, to point to the original linked-list; \textit{curr}, which points to the current node that should be iterated through or \textit{null}, if we are done iterating the list; \textit{index}, which indicates in what position the current node is in the nodes sequence; \textit{seen}, the sequence of values already iterated through; and \textit{to\_see}, the sequence of values still to be iterated.

\begin{minipage}{\linewidth}
\begin{lstlisting}[language=Java, caption={\it LinkedListIterator} class (part 1), label=\labellst{iteratorclass1}]
public final class LinkedListIterator {
  private /*@ spec_public @*/ LinkedList list;
  /*@ nullable @*/ private Node curr;
  //@ private ghost \bigint index;
  //@ private spec_public ghost \seq seen;
  //@ private spec_public ghost \seq to_see;
\end{lstlisting}
\end{minipage}

As usual, we define the iterator's footprint (List.~\reflst{iteratorclass2}). Such footprint is composed only by the iterator's fields (line 11). Notice that we do not include the list's footprint as part of the iterator's footprint. We also specify that the footprint itself does not change (line 9), and that the iterator's invariant depends on its footprint and on the list's footprint (line 10). The proof that nothing changes the footprint was done automatically. To prove that the invariant only depends on the iterator's and list's footprints, using KeY's default strategy was not helpful. The reason for this was that KeY was performing multiple ``cut'' tactics to attempt to close the proof for each possible value of \textit{list.size}. Because of this, we had to guide the proof. Nonetheless, it was mostly straightforward. We just had to use the ``observerDependency'' tactic which allowed us to establish that the iterator's invariant does not change in the presence of heap updates on locations that do not belong to its footprint or the list's footprint. Having the footprints public was also key to facilitate this result.

\begin{minipage}{\linewidth}
\begin{lstlisting}[language=Java, caption={\it LinkedListIterator} class (part 2), label=\labellst{iteratorclass2}, firstnumber=8]
  /*@ public model \locset footprint;
    @ accessible footprint: \nothing;
    @ accessible \inv: footprint, list.footprint;
    @ represents footprint = list, curr, index, seen, to_see;
    @*/
\end{lstlisting}
\end{minipage}

The iterator's invariant enforces properties over its fields and sequences (List.~\reflst{iteratorclass3}). First, we specify that \textit{index} is a value between zero and the number of values (line 15). This number may be equal to the number of values if and only if we have already iterated through all values. Then we enforce that the values in both sequences are not \textit{null} (line 16-19). This could be derived from the fact that the values in the list are not \textit{null}, but it is useful to keep this statement around. Following that, we define that the \textit{seen} sequence corresponds to the values already seen, from position zero (inclusive) to \textit{index} (exclusive) (line 20), and that \textit{to\_see} corresponds to the values to see, from position \textit{index} (inclusive) to the end (line 21). Since \textit{curr} points to the current node, we map it to position \textit{index} in the nodes sequence, or we specify that it is \textit{null}, when iteration is done (line 22). Finally, we keep the information that the list's invariant holds (line 23).

\begin{minipage}{\linewidth}
\begin{lstlisting}[language=Java, caption={\it LinkedListIterator} class (part 3), label=\labellst{iteratorclass3}, firstnumber=14]
  /*@ invariant
    @   0 <= index && index <= list.values.length &&
    @   (\forall \bigint i; 0 <= i < seen.length;
    @      (\exists FileReader f; f == seen[i] && f != null)) &&
    @   (\forall \bigint i; 0 <= i < to_see.length;
    @      (\exists FileReader f; f == to_see[i] && f != null)) &&
    @   seen == \seq_sub(list.values, 0, index) &&
    @   to_see == \seq_sub(list.values, index, list.values.length) &&
    @   (index < list.values.length ? curr == (Node)list.nodeList[index] : curr == null) &&
    @   \invariant_for(list);
    @*/
\end{lstlisting}
\end{minipage}

The constructor accepts two parameters: the list and the initial node. Its contract (List.~\reflst{iteratorclass4}) requires that the initial node be the \textit{head} of the list (line 29), and that the list's invariant holds. Then it guarantees that no location that is not fresh is modified (line 31). Finally, it ensures that, after initialization, this iterator is ``connected'' to the given list (line 33), that the sequence of seen values is empty (line 34), that the sequence of values to see corresponds to all values in the list (line 35), that the list's invariant still holds (line 36), and that the iterator's footprint is fresh (line 37). To verify the constructor, we had to guide the proof, mostly to establish the invariants of the iterator and list, since KeY was applying ``cut'' multiple times in the default strategy, as happened previously.

\begin{minipage}{\linewidth}
\begin{lstlisting}[language=Java, caption=Constructor's contract, label=\labellst{iteratorclass4}, firstnumber=26]
  /*@
    @ normal_behavior
    @ requires
    @   l.head == head &&
    @   \invariant_for(l);
    @ assignable \nothing;
    @ ensures
    @   list == l &&
    @   seen == \seq_empty &&
    @   to_see == l.values &&
    @   \invariant_for(l);
    @ ensures \fresh(footprint);
    @*/
\end{lstlisting}
\end{minipage}

The \textit{hasNext} method\footnote{\repolink{key/LinkedListIterator.java\#L47-L54}} returns \textit{true} is there are still values to iterate through. Its contract (List.~\reflst{iteratorclass5}) indicates that strictly nothing (i.e. no field of the object nor any other location, fresh or not) is assigned to (line 49), and that the boolean result is \textit{true} if and only if the sequence of values to see is not empty (line 50). This method was verified automatically with the default strategy.

\begin{minipage}{\linewidth}
\begin{lstlisting}[language=Java, caption={\it hasNext} method's contract, label=\labellst{iteratorclass5}, firstnumber=47]
  /*@
    @ normal_behavior
    @ assignable \strictly_nothing;
    @ ensures \result == (to_see.length > 0);
    @*/
\end{lstlisting}
\end{minipage}

The \textit{next} method\footnote{\repolink{key/LinkedListIterator.java\#L56-L75}} advances the iterator. Its contract (List.~\reflst{iteratorclass6}) requires that there are values to see (line 58). Then it specifies that the list's and iterator's footprints are the only accessible (i.e. read) heap locations (lines 59-60), and that only the iterator's footprint is assignable (line 61). When \textit{assignable} annotations are used without \textit{accessible} annotations, it is implicit that the assignable heap locations are the only accessible. In this instance, the list's footprint is not modified but it is read, so we need to use the \textit{assignable} annotation explicitly. Given that such an annotation is used, then we also need to say that the iterator's footprint is read. Following that, the method ensures that the returned value goes from the ``to see'' sequence to the ``seen'' sequence (lines 62-65). Finally, we state that the iterator is still ``connected'' to the same list as before (line 66), so that we do not lose this information.

\begin{minipage}{\linewidth}
\begin{lstlisting}[language=Java, caption={\it next} method's contract, label=\labellst{iteratorclass6}, firstnumber=56]
  /*@
    @ normal_behavior
    @ requires to_see.length > 0;
    @ accessible list.footprint;
    @ accessible footprint;
    @ assignable footprint;
    @ ensures
    @   seen == \seq_concat(\old(seen), \seq_singleton(\result)) &&
    @   to_see == \seq_sub(\old(to_see), 1, \old(to_see).length) &&
    @   \result == (FileReader)\old(to_see)[0];
    @ ensures list == \old(list);
    @*/
\end{lstlisting}
\end{minipage}

To verify the \textit{next} method, we had to prove that: 1. only the list's and iterator's footprints are accessible; 2. the post-condition holds after execution and that only the iterator's footprint is modified. Both proof requirements required a lot of work, likely because of the relation between \textit{index} and \textit{curr}, which was probably not obvious to KeY in the default strategy. Some examples of goals which required some effort to prove were: showing that the value of the current node was the first value in the ``to see'' sequence, proving that such a value was a file reader, and that the new sequences respected the invariant (after the current value was moved to the ``seen'' sequence).

\subsubsection*{Use example}

To exemplify the use of the linked-list with file readers, we instantiate a linked-list and add three file readers in the \textit{Init} state to it (lines 2-8 of List.~\reflst{main}). Then we pass the list to the \textit{useFiles} method (line 9) which iterates through all the files and executes their protocol until all of them reach the \textit{Closed} state. Note that this method may accept a list of any size.

\begin{minipage}{\linewidth}
\begin{lstlisting}[language=Java, caption=\textit{main} method, label=\labellst{main}]
public static void main(String[] args) {
  LinkedList list = new LinkedList();
  FileReader f1 = new FileReader();
  FileReader f2 = new FileReader();
  FileReader f3 = new FileReader();
  list.add(f1);
  list.add(f2);
  list.add(f3);
  useFiles(list);
}
\end{lstlisting}
\end{minipage}

The contract of the \textit{useFiles} method (List.~\reflst{usefilescontract}) requires that the list's invariant holds (line 4), that all the file readers' invariant holds, that they be in the \textit{Init} state (lines 5-6), and that all the file readers in the sequence of values of the list be distinct (lines 7-10). This method then ensures that all the file readers are in the \textit{Closed} state (lines 13-14). Only the file readers are modified by this method (line 11).

\begin{minipage}{\linewidth}
\begin{lstlisting}[language=Java, caption=\textit{useFiles} contract, label=\labellst{usefilescontract}, breaklines=true]
/*@
  @ normal_behavior
  @ requires
  @   \invariant_for(list) &&
  @   (\forall \bigint i; 0 <= i < list.values.length;
  @       (\exists FileReader f; f == list.values[i] && f != null && \invariant_for(f) && f.state == FileReader.STATE_INIT)) &&
  @   (\forall \bigint i; 0 <= i < list.values.length;
  @     (\forall \bigint j; 0 <= j < list.values.length;
  @       (FileReader)list.values[i] == (FileReader)list.values[j] ==> i == j
  @   ));
  @ assignable (\infinite_union \bigint i; 0 <= i < list.values.length; ((FileReader)list.values[i]).footprint);
  @ ensures
  @   (\forall \bigint i; 0 <= i < list.values.length;
  @     (\exists FileReader f; f == list.values[i] && f != null && \invariant_for(f) && f.state == FileReader.STATE_CLOSED));
  @*/
public static void useFiles(LinkedList list) { ... }
\end{lstlisting}
\end{minipage}

The verification of the \textit{main} method (List.~\reflst{main}) required a lot of interactivity, namely to show that the pre-condition of the \textit{useFiles} method was satisfied. Showing that the list's invariant was straightforward. To prove that the file readers in the values sequence were all distinct, we used the Z3 solver. Proving that all the file readers were in the initial state was the most laborious task. The technique we used was to ``cut'' on all the possible index values (i.e. 0, 1, 2), and prove for each file reader independently that it was in fact in the initial state. This also requires showing that the \textit{add} operations on the list do not modify the file readers, which in turn requires showing that the list's footprint is disjoint from the file readers' one. Having the footprints public was also important here.

The initial code for the \textit{useFiles} method was based on using an iterator to iterate through each file using a \textit{while} loop (List.~\reflst{usefiles1}). Unfortunately, it was very difficult to verify such implementation, because symbolic execution would open all the heap accesses in the hypothesis and goal, which resulted in several \textit{if} statements that could not be simplified, making it very hard to read what was available. These statements relied on showing that, for example, the iterator's footprint was disjoint from the resulted file reader. Enabling the automatic opening of class invariant would help with this but it would also be unnecessary in several cases. To facilitate the proofs, we decided to instead implement a recursive version, although this would not be the most efficient or natural implementation.\footnote{\repolink{key/Main.java\#L18-L36}}

\begin{minipage}{\linewidth}
\begin{lstlisting}[language=Java, caption=\textit{useFiles}: initial implementation, label=\labellst{usefiles1}]
LinkedListIterator it = list.iterator();
// loop invariant omitted
while (it.hasNext()) {
  FileReader f = it.next();
  use(f);
}
\end{lstlisting}
\end{minipage}

The recursive version of the \textit{useFiles} method (List.~\reflst{usefiles2}) starts by creating an iterator for the linked-list (line 1) and then passing it to an helper method called \textit{recursive} (line 2), which iterates for each file reader.\footnote{\repolink{key/Main.java\#L38-L77}} The most difficult part was to show that the sequence of values was equal to the ``seen'' sequence of file readers, and that the original values sequence was not modified. We could show this because \textit{recursive} returns leaving the ``to see'' sequence empty.

\begin{minipage}{\linewidth}
\begin{lstlisting}[language=Java, caption=\textit{useFiles}: recursive version, label=\labellst{usefiles2}]
LinkedListIterator it = list.iterator();
recursive(list, it);
\end{lstlisting}
\end{minipage}

The \textit{recursive} method's contract is very verbose (List.~\reflst{recursivemethod}). It requires that the iterator's and list's invariants hold (lines 4-5), that both are ``connected'' (line 6), that the files ``seen'' are \textit{Closed} (lines 7-8), that the files ``to see'' are in the \textit{Init} state (lines 9-10), and that the file readers in both sequences are distinct (lines 11-18). It then ensures that only the iterator and files ``to see'' are modified (lines 19-20), and that the pre-conditions are preserved (lines 22-31) except that the ``to see'' sequence returns empty (line 32). We also provide a decreasing argument (line 21), the size of the ``to see'' sequence, using the \textit{measured\_by} annotation, to ensure termination and allow us to use its own contract when checking the recursive call. Omitting it or using the \textit{decreases} annotation (which does nothing in methods) prevents us from doing this. This method calls \textit{hasNext} to check for file readers to read (line 35). If there are, we extract one (line 36) and call the \textit{use} method, previously presented (List.~\reflst{fileusage}), to read the file and close it (line 37). Then we perform a recursive call (line 38).

\begin{minipage}{\linewidth}
\begin{lstlisting}[language=Java, caption=\textit{recursive} method, label=\labellst{recursivemethod}, breaklines=true]
/*@
  @ normal_behavior
  @ requires
  @   \invariant_for(it) &&
  @   \invariant_for(list) &&
  @   it.list == list &&
  @   (\forall \bigint i; 0 <= i < it.seen.length;
  @     (\exists FileReader f; f == it.seen[i] && f != null && f.state == FileReader.STATE_CLOSED && \invariant_for(f))) &&
  @   (\forall \bigint i; 0 <= i < it.to_see.length;
  @     (\exists FileReader f; f == it.to_see[i] && f != null && f.state == FileReader.STATE_INIT && \invariant_for(f))) &&
  @   (\forall \bigint i; 0 <= i < it.seen.length;
  @     (\forall \bigint j; 0 <= j < it.seen.length;
  @       (FileReader)it.seen[i] == (FileReader)it.seen[j] ==> i == j
  @   )) &&
  @   (\forall \bigint i; 0 <= i < it.to_see.length;
  @     (\forall \bigint j; 0 <= j < it.to_see.length;
  @       (FileReader)it.to_see[i] == (FileReader)it.to_see[j] ==> i == j
  @   ));
  @ assignable it.footprint;
  @ assignable (\infinite_union \bigint i; 0 <= i < it.to_see.length; ((FileReader)it.to_see[i]).footprint);
  @ measured_by it.to_see.length;
  @ ensures
  @   \invariant_for(it) &&
  @   \invariant_for(list) &&
  @   it.list == list &&
  @   (\forall \bigint i; 0 <= i < it.seen.length;
  @     (\exists FileReader f; f == it.seen[i] && f != null && f.state == FileReader.STATE_CLOSED && \invariant_for(f))) &&
  @   (\forall \bigint i; 0 <= i < it.seen.length;
  @     (\forall \bigint j; 0 <= j < it.seen.length;
  @       (FileReader)it.seen[i] == (FileReader)it.seen[j] ==> i == j
  @   )) &&
  @   it.to_see == \seq_empty;
  @*/
private static void recursive(LinkedList list, LinkedListIterator it) {
  if (it.hasNext()) {
    FileReader f = it.next();
    use(f);
    recursive(list, it);
  }
}
\end{lstlisting}
\end{minipage}

To verify this method, we guided the proof, including the steps of symbolic execution, for mainly two reasons: to control exactly what would be opened and to show early one that \textit{hasNext} returning \textit{true} means that the ``to see'' sequence is not empty. Of course, the goals that could be closed automatically with KeY or Z3, we did not prove manually.

The most challenging part of verifying this method was (again) related with showing that some parts of the heap were not modified (for example, that the \textit{use} call in line 37 did not modify the iterator nor the sequences), which relied on showing that certain footprints were disjoint. It was also difficult to show that only the file readers in the initially received ``to see'' sequence were modified. The solution was to open two goals: one where a given field was considered to be in the infinite union described in line 20, and another where that was not the case. Finally, establishing the pre-condition of the recursive call required some work because of the file reader that was moved from the ``to see'' sequence to the ``seen'' one.

\subsubsection*{Protocol completion}

To ensure that all file readers reach the end of their protocol, we define a contract for the \textit{main} method such that for all file readers created at some point in the program, they are in the \textit{Closed} state (line 3 of List.~\reflst{mainwithcompletion}). Since KeY is not aware of the objects created or not before, we define as pre-condition that no file readers exist when \textit{main} is called (line 2). This works because quantifiers in KeY only reason over objects in the heap. Given that \textit{main} is the first method called in a Java program, this requirement is actually an assumption. In this example, the \textit{use} calls return with the file readers they are given in the \textit{Closed} state (lines 10-12).

\begin{minipage}{\linewidth}
\begin{lstlisting}[language=Java, caption=\textit{main} method with protocol completion, label=\labellst{mainwithcompletion}]
/*@
  @ normal_behavior
  @ requires !(\exists FileReader f; true);
  @ ensures (\forall FileReader f; f.state == FileReader.STATE_CLOSED);
  @*/
public static void main(String[] args) {
  FileReader f1 = new FileReader();
  FileReader f2 = new FileReader();
  FileReader f3 = new FileReader();
  use(f1);
  use(f2);
  use(f3);
}
\end{lstlisting}
\end{minipage}

Because KeY's verification procedure is modular, and since we are not interested in unfolding any method's implementation, instead just relying on their contracts, we actually have to specify that no fresh file readers are created in all other methods. This is necessary because it would be possible for methods to create file readers only available in the scope of their execution, which would exist in the heap as created objects, but for which we would know nothing about. This post-condition, written as \texttt{!(\textbackslash exists FileReader f; \textbackslash fresh(f))}, was added in all methods of the file reader and in the \textit{use} method. In the file reader's constructor, we also had to say that \textit{f} was different from \textit{this}, since the newly created reference is fresh.

All the adapted file reader's methods were verified with the default strategy. The \textit{main} and \textit{use} methods were verified by first running the default strategy following by Z3 to close all the remaining goals. This example with protocol completion is also available online.\footnote{\repolink{key/protocol-completion}}

Since KeY is based on first-order dynamic logic, it is very simple to use quantifiers to specify that all file readers should have their protocol completed, independently from where they were created during the execution of \textit{main}. Although this seems very powerful, we have to adapt all other methods to specify that no new file readers are created inside.

\subsubsection*{Assessment}

The use of JML for specifications, a language for formally specifying behavior of Java code, used by various tools, reduces the learning curve for those that already know JML. Furthermore, KeY supports the greatest number of Java features, allowing one to verify real programs considering the actual Java runtime semantics. Nonetheless, generics are not supported, although there is an automated tool to remove generics from Java programs, which can then be verified with KeY.\footnote{\url{https://www.key-project.org/docs/user/RemoveGenerics/}} Because of its focus on Java, KeY is not overly suitable for the verification of algorithms that require abstracting away from the code~\cite{DBLP:journals/sttt/BrunsMU15}.

One important aspect that makes KeY stand out from other tools is the support for interactivity, which allows the programmer to guide the proofs. This is an aspect that we missed when experimenting with other tools. Additionally, KeY provides useful macros and a high degree of automation as well as support for SMT solvers, such as Z3. We used Z3 often to more quickly close provable goals, specially those involving universal quantifiers.

Since KeY's core is based on first-order dynamic logic, one can express heap-dependent expressions. In fact, the heap is an object in the logic that is mentioned explicitly. Although this allows for much expressiveness, it often becomes very difficult to verify programs. For example, when attempting to prove that the footprints of two objects were disjoint, we also had to account for possible changes in the footprints themselves, which became very cumbersome. To do this, we had to make the footprints public so that they could be opened in proofs, otherwise we are not sure if we would have been able to finish the proofs. We think that a variant of KeY that would instead use separation logic, to abstract away the heaps and the notion of disjointness, would be very helpful and improve readability. Additionally, it would avoid the need to explicitly specify that, for example, the values in a sequence are all distinct, thus reducing the specifications' verbosity. Because we had to be explicitly aware of issues regarding heap disjointness, it was often the case that we had to repeat proofs because we realized that the specification was incorrect (e.g. wrong \textit{assignable} clause) or insufficient to prove something (e.g. missing the \textit{assignable} clause).

Nonetheless, there are probably other alternatives to separation logic that would help in solving the aforementioned issues. In a private conversation with KeY's group leaders~\cite{private_conversation_reiner_hahnle}, they point out that the connectives of separation logic ``would get in the way of automation'', a crucial feature of KeY. For example, it seems that ``the heap separation rule tends to split proofs too early''.
Furthermore, the problems we encountered can be summarized in two main points: (1) insufficient abstract specification primitives; (2) inability to prove heap and functional properties separately, in a modular fashion. KeY's team is aware of these issues and will address them in the future (at the time of writing).

In the context of typestates, checking for protocol completion is crucial to ensure that necessary method calls are not forgotten and that resources are freed, thus avoiding memory leaks. Although such concept is not built-in in separation logic, this logic makes it immediately clear which heap chunks are accessed and potentially modified. Together with resource leaking prevention (except for objects with completed protocol), one could ensure protocol completion without the need for adding extraneous post-conditions to all methods. This would be another reason why we believe separation logic would be preferable over first-order dynamic logic, but it is possible there are other alternatives.

We enjoyed the user experience and appreciated that KeY comes with examples to experiment with. Nonetheless, we believe there is room for improvement. For instance, when we first tried to use KeY, we right-clicked different expressions, which became highlighted as we moved the mouse over them. But the menu that opens up does not provide all the available options. Only later we realized that left-clicking shows a menu with more options.
We also think that ``Ctrl+Z'' should return the proof to the last interactive step, and not the last step regardless if it was an automatic step part of a macro or if it was an interactive one.
Additionally, we think symbolic execution should avoid opening heap accesses and other kinds of expressions by default, to facilitate the reading process. We also noticed that often several equalities were generated, some mentioning names such as \textit{LinkedList\_values\_5$<<$selectSK$>>$}, which are difficult to read. It would be useful to have a macro that would perform substitutions, like \textit{subst} in Coq.\footnote{\url{https://coq.inria.fr/refman/proofs/writing-proofs/equality.html\#coq:tacn.subst}}
Furthermore, we were surprised that expressions such as \textit{list.length} could be parsed, while expressions like \textit{list.length@heap\_afterAdd} could not. These two are both the prettified version of their more verbose counterpart, but one is recognized and the other is not. We often had to turn off ``Use Pretty Syntax'' to copy expressions in their ``real'' form and use them in tactics.

Finally, the support for saving and reusing parts of proofs was very useful to us. We wonder if it would be possible to improve such reutilization of proofs in the presence of slight changes to specifications. It often occurred that a whole proof tree would need to be discarded because some part of a specification or code was modified. When it is necessary to guide a big part of a proof, that kind of situation can be very frustrating to the user. A questionnaire evaluating the usability of KeY also makes this point, among others we mentioned, like the need to improve readability~\cite{DBLP:conf/cade/BeckertG12}.

\section{Conclusions}
\label{sec:conclusions}

In this survey, we reviewed four tools for verifying Java code. In particular, we evaluated their ability to check the correct use of objects with protocol, even when these were shared in collections, if they were able to guarantee protocol completion, and the programmer's effort in making the code acceptable to each.

Both VeriFast and VerCors are based on separation logic and support logical predicates, which allow for rich and expressive specifications. Although the expressiveness is useful, deductive reasoning via lemmas is often required. This is very demanding and can be a barrier to less experienced users. We believe improved interactive experiences for programmers are key to make these tools more approachable. In our experience with these two tools, we spent more time and lines (between about 100 and 160) to prove results than to implement the code. Furthermore, fractional permissions only allow for read-only access when data is shared. In consequence, either locks are required to mutate shared data (even in single-threaded code, where they are not really necessary, resulting in inefficient code), or a complex specification workaround is needed.

Plural is different from these tools in two major ways: it does not support logical predicates and so, specifications are less expressive in that regard, which prevented us from implementing a linked-list, but the use of access permissions allows for more kinds of sharing. However, it is still true that accesses to thread-local shared data might require the use of locks. The annotation effort was minimal but that may be the consequence of having less expressive specifications.

KeY stands out because of its support for interactivity as well as useful macros, automation, and the ability to save and reuse proofs. Nonetheless, the fact that heaps are mentioned explicitly in logical assertions makes it sometimes difficult to read the hypothesis and the proof goals. Additionally, it is often the case that one needs to show that certain footprints are disjoint, and because the proof process may be tedious, this can drive users away. Even though it is true that KeY can automate fully some proofs, that depends on finding the right specifications and proof search settings, which is not easy for beginners. More abstract specification primitives, and the ability to prove heap and functional properties separately, are crucial features to improve the user experience, in terms of both readability and ease of proving results.

Furthermore, as we have noted, protocol completion is crucial to ensure that necessary method calls are not forgotten and that resources are freed. Unfortunately, none of these tools directly supports static guarantees of such. Although tools like VeriFast and VerCors allow for workarounds, we believe such guarantee should be provided directly by the type system. This could be supported by ensuring that no permission to an object is ``dropped'' unless it is in the final state. An automated static analysis based on a type system checking for protocol compliance and completion is essential because it provides these guarantees without the need for much input from the programmer.

Given the annotation effort often required, either to prove results, or to define the behavior of programs, as another study also highlights~\cite{DBLP:conf/icse-formalise/LathouwersH22}, it is imperative that easier ways to specify protocols be provided. Additionally, better techniques should be available to reason about the permissions to memory locations, in a way that does not force the programmer to spend time proving that a certain permission is available, time that could be used to focus on ensuring the correctness of the program. In conclusion, this study motivates the need for a language that statically provides the guarantees of protocol compliance and completion in the presence of several patterns of object sharing.

\clearpage

\bibliographystyle{elsarticle-num-names}
\bibliography{bibliography.bib}

\end{document}